\documentclass[a4paper,11pt]{article}

\usepackage{amsmath,amsfonts,amssymb,caption,graphicx} % ,natbib}
\usepackage{tikz}
\usetikzlibrary{trees}
\newcommand{\Y}{{\rm Y}}

\newcommand{\Xp}{\mbox{\boldmath $X$}}

\newcommand{\Yp}{\mbox{\boldmath $Y$}}

\newcommand{\weakly}{\mbox{$ \;\stackrel{\cal D}{\longrightarrow}\; $}}

\newcommand{\RL}{{\mathbb R}}

\newcommand{\IND}{{\mathbb I}}

\newcommand{\VAR}{\mbox{\rm Var}}

\def\ba{\begin{align}}
\def\ea{\end{align}}
\def\ban{\begin{align*}}
\def\ean{\end{align*}}

\def\be{\begin{eqnarray}}
\def\ee{\end{eqnarray}}
\def\ben{\begin{eqnarray*}}
\def\een{\end{eqnarray*}}

\def\bqq{\begin{equation}}
\def\eqq{\end{equation}}
\def\bqqn{\begin{equation*}}
\def\eqqn{\end{equation*}}

\def\elabel#1{\label{e:#1}}

\def\sq{$\Box$}

\def\qed{\ifmmode\sq\else{\unskip\nobreak\hfil
\penalty50\hskip1em\null\nobreak\hfil\sq
\parfillskip=0pt\finalhyphendemerits=0\endgraf}\fi\par\medbreak}

\newsavebox{\junk}
\savebox{\junk}[1.6mm]{\hbox{$|\!|\!|$}}

\def\til={{\widetilde =}}

\def\clK{{\cal K}}
\def\clL{{\cal L}}

 \def\eq#1/{(\ref{#1})}

\newtheorem{theorem}{Theorem}[section]
\newtheorem{corollary}[theorem]{Corollary}
\newtheorem{proposition}[theorem]{Proposition}

\def\eq#1/{(\ref{e:#1})}

\newcommand{\beqn}[1]{\notes{#1}%
\begin{eqnarray} \elabel{#1}}

\newcommand{\eeqn}{\end{eqnarray} }

\newcommand{\beq}[1]{\notes{#1}%
\begin{equation}\elabel{#1}}

\newcommand{\eeq}{\end{equation}} 

\def\bdes{\begin{description}}
\def\edes{\end{description}}

\def\notes#1{}

\begin{document}

\title{\vspace{-1.5cm}%
Estimating the Directed Information 
and Testing for Causality\footnotetext{This work 
was supported by the European Union and Greek National
Funds through the Operational Program Education and Lifelong Learning of
the National Strategic Reference Framework through the Research Funding
Program Thales-Investing in Knowledge Society through the European Social
Fund.}
	\footnotetext{Preliminary versions 
	of some of the results in this paper
	were presented at the Eighth Workshop on Information Theoretic 
	Methods in Science and Engineering,
	24-26 June 2015, Copenhagen, Denmark.}
}

\author
{
        Ioannis Kontoyiannis, {\sl Fellow, IEEE}
    \thanks{Department of Informatics,
        Athens University of Economics and Business,
        Patission 76, Athens 10434, Greece.
                Email: {\tt yiannis@aueb.gr}.
        }
% \and
	% Athina Panotopoulou
    % \thanks{Department of Informatics,
        % Athens University of Economics and Business,
        % Patission 76, Athens 10434, Greece.
                % Email: {\tt atpanot@gmail.com}.
	% }
\and
	Maria Skoularidou
    \thanks{Department of Informatics,
        Athens University of Economics and Business,
        Patission 76, Athens 10434, Greece.
		Email: {\tt m.skoularidou@gmail.com}.
	}
}

\date{\today}

\maketitle

\begin{abstract}

The problem of estimating the directed 
information rate between two discrete processes 
$\{X_n\}$ and $\{Y_n\}$ via the plug-in
(or maximum-likelihood) estimator is considered.
When the joint process $\{(X_n,Y_n)\}$
is a Markov chain of a given memory length,
the plug-in estimator is shown to be
asymptotically Gaussian and to 
converge at the optimal rate 
$O(1/\sqrt{n})$ under appropriate
conditions; this is the first 
estimator that has been shown to 
achieve this rate.
An important connection is drawn
between the problem of estimating
the directed information rate
and that of performing a hypothesis test 
for the presence of causal influence 
between the two processes. 
Under fairly general conditions,
the null hypothesis, which corresponds 
to the absence of causal influence, 
is equivalent to the 
requirement that the directed
information rate be equal to zero.
In that case a finer result is
established, showing that the 
plug-in converges 
at the faster rate $O(1/n)$
and that it is asymptotically
$\chi^2$-distributed.
This is proved by showing
that this estimator is 
equal to (a scalar multiple 
of) the classical likelihood ratio
statistic for the above hypothesis 
test. Finally it is noted that
these results
facilitate the design of 
an actual likelihood ratio
test for the presence or
absence of causal influence.
\end{abstract}


% \footnotetext{This work was supported in part by
	% the European Union (European Social Fund 
	% - ESF) and Greek national funds through the Operational Program 
	% ``Education and Lifelong Learning'' of the National Strategic 
	% Reference Framework (NSRF) Research Funding Program 
	% ``Thales - Investing in knowledge society through the 
	% European Social Fund.''}

\bigskip

%%%%%%%%%%%%%
%%%%%%%%%%%%%
%%%%%%%%%%%%%
%%%%%%%%%%%%%

\thispagestyle{empty}
%\clearpage
\setcounter{page}{1}

\newpage

\section{Introduction}
\label{s:intro}
%%%%%%%%%%%%%%%%%%%%%%%%%%%%%%%%%%%%%%%%%%%%%%%%%%%%%%%%%%%%%%%%%%%%%%

{\bf Hypothesis testing and mutual information.}
%%%%%%%%%%%%%%%%%%%%%%%%%%%%%%%%%%%%%%%%%%%%%%%%%%%%%%%%%%%%%%%%%%%%%%
One of the most prevalent statistical tools
used universally across the sciences today, is the
$\chi^2$ test for independence. Suppose we have 
independent and identically distributed data pairs
$(X_1,Y_1),$ $(X_2,Y_2),\ldots,(X_n,Y_n)$ and we wish
to test whether the $X$ and $Y$ variables are independent
or not. Assuming both sets of variables take on
finitely many values,
we can compare the joint empirical distribution 
$\hat{P}_{XY,n}(a,b)$ with the product of the
empirical marginals $\hat{P}_{X,n}(a)\hat{P}_{Y,n}(b)$;
as usual, $\hat{P}_{XY,n}(a,b)$ denotes the proportion
of times the pair $(a,b)$ appears in the whole sample,
and similarly for the marginals. Pearson's $\chi^2$ test
dictates that we compute the (normalized)
$\chi^2$ distance between
these two distributions,
\be
\bar{\chi}_n^2=n\sum_{a,b}\frac{\big[\hat{P}_{XY,n}(a,b)
-\hat{P}_{X,n}(a)\hat{P}_{Y,n}(b)\big]^2}
{
\hat{P}_{X,n}(a)\hat{P}_{Y,n}(b)
}.
\label{eq:test1}
\ee
If the data are indeed independent then 
the distribution of the statistic $\bar{\chi}_n^2$,
for large sample sizes $n$, is approximately
$\chi^2$ with $(m-1)(\ell-1)$ degrees of freedom,
where $m,\ell$ are the sizes of the alphabets
of $X$ and $Y$, respectively. Therefore, we can
compute the probability of observing 
a value greater than or equal to
$\bar{\chi}^2_n$ under this distribution, 
and if this probability is appropriately small 
then we can reject the independence hypothesis. 

Another classical test, more closely related to
information theoretic ideas, is the likelihood
ratio test, based on the statistic,
$$\Delta_n=2\log\left(\frac
{\prod_{i=1}^n\hat{P}_{XY,n}(X_i,Y_i)}
{\prod_{i=1}^n\hat{P}_{X,n}(X_i)\hat{P}_{Y,n}(Y_i)}
\right).
$$
This has the exact same asymptotic distribution
as $\bar{\chi}^2_n$, and an 
analogous test can be performed. 
In fact, the $\bar{\chi}^2_n$ statistic can 
be viewed (and sometimes its use is thus
justified) as a quadratic approximation to the
nonlinear statistic $\Delta_n$.
A more important observation for 
our purposes is that, after simple algebra, 
the likelihood ratio test statistic
can exactly be expressed as a mutual information,
\be
\Delta_n=2nI(\hat{X};\hat{Y})
=2nD\big(\hat{P}_{XY,n}\|\hat{P}_{X,n}\hat{P}_{Y,n}\big),
\label{eq:test2}
\ee
where the random variables $\hat{X},\hat{Y}$
have distribution $\hat{P}_{XY,n}$.
Therefore, instead of the $\chi^2$ distance used 
in~(\ref{eq:test1}), 
the likelihood ratio test statistic~(\ref{eq:test2}) 
examines
the (normalized) relative entropy distance between 
$\hat{P}_{XY,n}$ and 
$\hat{P}_{X,n}\hat{P}_{Y,n}$.
And yet another way to interpret $\Delta_n$ is as
the ``plug-in'' estimate of the mutual information
$I(X_1;Y_1)$ of the data, using their empirical 
distribution. 

The asymptotic distribution of $\Delta_n$
has been re-derived several
times historically. In its general form it goes
back to the classical result of 
Wilks \cite{wilks:38}, see also
the texts \cite{lehmann-romano:book,schervish:book};
in more recent years it has also reappeared 
in an information-theoretic context,
see, e.g.,
\cite{hagenauer-et-al:04,hagenauer-et-al:05,spaKetal:07}.

\bigskip

\noindent
{\bf Estimating directed information and causality testing.}
%%%%%%%%%%%%%%%%%%%%%%%%%%%%%%%%%%%%%%%%%%%%%%%%%%%%%%%%%%%%%%%%%%%%%%
This work examines
the problem of estimating a different
information-theoretic functional: 
If $\Xp=\{X_n\}$ and $\Yp=\{Y_n\}$ are two finite-valued
random process, then the {\em directed information}
$I(X_1^n\to Y_1^n)$ between $X_1^n=(X_1,X_2,\ldots,X_n)$
and $Y_1^n=(Y_1,Y_2,\ldots,Y_n)$ is defined as,
$$I(X_1^n\to Y_1^n)=H(Y_1^n)-\sum_{i=1}^n H(Y_i|Y_1^{i-1},X_1^i),$$
and the {\em directed information rate} between
$\Xp$ and $\Yp$ is,
\be
I(\Xp\to\Yp)=\lim_{n\to\infty}
\frac{1}{n} I(X_1^n\to Y_1^n),
\label{eq:DIdefn}
\ee
whenever the limit exists; precise definitions
are given in Sections~\ref{s:I} and~\ref{s:DI}.
Directed information was introduced by 
Massey \cite{massey:90}, building on earlier 
work by Marko \cite{marko:73}, in order to provide 
capacity characterizations for channels 
with causal feedback. Subsequent work in this
direction includes
\cite{kramer:phd,tatikonda:phd,kim:08,permuter-et-al:09,shrader:09,dabora:10},
and the dual problem of lossy data compression with
feedforward is treated, e.g., in \cite{VP-DMI:07}.
Additional areas where directed information
plays an important role include, among others,
distributed and causal data compression 
and hypothesis testing
\cite{gunduz-erkip:07,weissman-et-al:11},
network communications and control
\cite{kramer:14pre,silva-et-al:pre},
dynamically switching networks
\cite{haskovec:15},
sensor networks \cite{rahimian:13,zaidi:14},
and causal estimation \cite{weissman-et-al:13};
see also the references in the above works.

Here we consider the problem of estimating
the directed information rate, by tracing the
path described above in connection with the 
mutual information in the reverse direction.
Assuming that the joint process
$\{(X_n,Y_n)\}$ is a Markov chain of memory length
$k\geq 1$, we begin by recalling that under 
fairly general conditions
the limit~(\ref{eq:DIdefn}) can be expressed 
in more manageable form. For example,
in Proposition~\ref{prop:markovDI} we note that 
when $\{Y_n\}$ itself is also
a Markov chain of order no greater than
$k$, the directed information rate is equal 
to the conditional mutual information
$I(\bar{Y}_0;\bar{X}_{-k}^0|\bar{Y}_{-k}^{-1})$, where
$\{(\bar{X}_n,\bar{Y}_n)\}$ denotes the stationary
version of the chain.

\bigskip

\noindent
{\bf Main results.}
%%%%%%%%%%%%%%%%%%%%%%%%%%%%%%%%%%%%%%%%%%%%%%%%%%%%%%%%%%%%%%%%%%%%%%
In Section~\ref{s:DI-pi} we define 
what is probably the simplest estimator 
of $I(\Xp\to\Yp)$:
Interpreting the conditional mutual information
$I(\bar{Y}_0;\bar{X}_{-k}^0|\bar{Y}_{-k}^{-1})$
as a functional of the $(k+1)$-dimensional distribution
of $(\bar{X}_{-k}^0,\bar{Y}_{-k}^0)$, we define
the plug-in estimator $\hat{I}_n^{(k)}(\Xp\to\Yp)$
as the same functional of the 
corresponding empirical distribution.
If the original 
chain is ergodic, then it is easy to see that
this estimator is consistent with probability one. 
Our main results,
stated in Theorems~\ref{thm:DI-markovY}
and~\ref{thm:DI-markov}, give finer 
information for this convergence;
cf.~(\ref{eq:CLT}) and~(\ref{eq:chisqI})
below.

On the one hand, if
$I(\Xp\to\Yp)>0$ we
show that, under appropriate conditions,
the plug-in estimator is approximately 
Gaussian for large $n$,
\be
\hat{I}_n^{(k)}(\Xp\to\Yp)\approx 
N\left(I(\Xp\to\Yp),\frac{\sigma^2}{n}\right),
\label{eq:CLT}
\ee
where the variance $\sigma^2$
is identified in 
Theorem~\ref{thm:DI-markovY}.
Therefore, 
the plug-in
estimator converges at a rate $O(1/\sqrt{n})$
in probability. In fact, 
Corollary~\ref{cor:DI-LIL}
shows that the same rate holds in
$L^1$, and in view of 
\cite[Proposition~3]{weissman-et-al:di}
this implies that the plug-in estimator
is optimal in that it
converges at the fastest possible rate.
Moreover, in 
Corollary~\ref{cor:DI-LIL} we also
establish the almost-sure convergence
rate of the plug-in estimator, under
fairly general conditions.

On the other hand we note that $I(\Xp\to\Yp)=0$
if and only if a certain conditional 
independence property holds, which can be 
interpreted as the absence of causal influence 
from $\Xp$ to $\Yp$: Roughly speaking
$I(\Xp\to\Yp)$ is zero if and only 
if each $Y_i$, given the past values of the 
$\Yp$ process, is conditionally independent of 
the values of the $\Xp$ process up to time $i$. 
In fact, one of the main
contributions of this work is the 
identification of two related problems:
(1.)~understanding
the asymptotics of the plug-in 
estimator $\hat{I}_n^{(k)}(\Xp\to\Yp)$;
and~(2.)~analyzing the likelihood ratio test 
for the above causality hypothesis. 

Intuitively, determining whether the estimated
directed information $\hat{I}_n^{(k)}(\Xp\to\Yp)$
is significantly close to zero or not,
is related to testing the 
hypothesis that the above conditional 
independence relationship holds.
Formally, as we show in 
Proposition~\ref{prop:DILRT}, 
the (normalized) likelihood ratio statistic
for this test is {\em exactly} equal to
the plug-in
estimator $\hat{I}_n^{(k)}(\Xp\to\Yp)$.
This connection is described in detail
in Section~\ref{s:DI-HT}. Apart from being
intellectually satisfying, it also allows us
to derive the exact asymptotic distribution 
of $\hat{I}_n^{(k)}(\Xp\to\Yp)$.

Indeed,
% in Theorems~\ref{thm:DI-markovY} and~\ref{thm:DI-markov} 
we show that,
under appropriate conditions, 
if the directed information rate is zero,
then a finer result than~(\ref{eq:CLT})
can be established
for the plug-in estimator;
for large $n$ we have,
\be
\hat{I}_n^{(k)}(\Xp\to\Yp)\approx 
\frac{1}{n}{\rm X}^2(m,\ell,k),
\label{eq:chisqI}
\ee
where ${\rm X}^2(m,\ell,k)$ is an appropriate
$\chi^2$ distribution that only
depends on the sizes $m,\ell$ of the alphabets
of $X$ and $Y$ and on the memory size $k$.
In other words, under the null hypothesis
(which asserts the presence of the conditional
independence being tested), the plug-in
estimator converges at a faster rate $O(1/n)$,
and the distribution to which it converges
only depends on the three known parameters
$m,\ell$ and $k$. Therefore, a likelihood
ratio test for this type of causal influence,
can again be performed as before.

In Section~\ref{s:I} we consider
the simpler problem of estimating the
mutual information rate
$\lim_n I(X_n;X_0^{n-1})$ of a Markov chain
$\{X_n\}$. The results presented there can
be viewed as preliminary versions of the analogous
results for the directed information given
in Section~\ref{s:DI}. For clarity of exposition,
all proofs are collected in the Appendix.

\bigskip

\noindent
{\bf Earlier work.}
%%%%%%%%%%%%%%%%%%%%%%%%%%%%%%%%%%%%%%%%%%%%%%%%%%%%%%%%%%%%%%%%%%%%%%
The connection between the problem of 
identifying causal relationships and that of
testing for conditional independence has a long history. 
Perhaps the most prominent example is the 
Granger causality test \cite{granger:69},
which uses an autoregressive model 
(later extended in several directions,
most notably to generalized linear models), 
within which the conditional independence 
hypothesis described above is tested.
The connection of this test
with directed information
has previously been explored in several directions;
see \cite{amblard:12} for a comprehensive review.

Several different approaches to
the problem of directed information
estimation have appeared in the
literature in recent years. 
Rao {\sl et al.} \cite{rao-etal:08} use 
Miller's
\cite{miller:03} 
differential entropy estimators 
in order to estimate (the continuous analog of)
directed information, in order 
to identify causal influence in networks 
of genes. In the context of neuroscience,
Quinn {\sl et al.}
\cite{neurspik} use parametric estimation
based on generalized linear models to
estimate directed information, in order to
detect the presence of direct 
or indirect influence in neuronal networks. 
And in the subsequent work~\cite{quinn:13},
a near-optimal rate of convergence 
$O(n^{-1/2+\epsilon})$ is established 
for the plug-in estimator.

In terms 
of the present development,
the most interesting work
is the recent paper by 
Jiao {\sl et al.} \cite{weissman-et-al:di}.
There, several new estimators for the
directed information rate are introduced
and they are shown to be consistent under
very general conditions
For some of these
estimators, particularly those based on
the context tree weighting
algorithm \cite{willems-shtarkov-tjalkens:95},
detailed convergence bounds are also obtained.
It is worth noting that our convergence results 
are obtained under conditions essentially
identical to (though slightly 
weaker than) those required for 
the bounds in~\cite{weissman-et-al:di}.
% Compared to the estimators of 
% \cite{weissman-et-al:di}, the plug-in suffers
% two well-known drawbacks. 
% Its use is restricted to Markovian data,
% and it is computationally
% ineffective for large alphabet sizes and
% long memory processes -- although in future
% work we plan to argue that it is still very
% effective in many nontrivial circumstances.
But using the plug-in also facilitates
the connection with hypothesis testing
developed here, and makes it possible
to obtain, instead of convergence bounds,
accurate and sometimes 
exact asymptotics as described above.

Finally, in a broader context, 
we note 
the $L^1$ and $L^2$ convergence results
presented for mutual information
in the very recent work \cite{han:15}.
Although the problems treated there are mostly
in the case of independent observations,
they provide a general minimax framework
for examining the asymptotic optimality
of different estimators.

\bigskip

\noindent
{\bf Different approaches.}
%%%%%%%%%%%%%%%%%%%%%%%%%%%%%%%%%%%%%%%%%%%%%%%%%%%%%%%%%%%%%%%%%%%%%%
The problem of estimating directed information 
via the plug-in can and has arisen as a natural 
question within several different areas.
Motivated by applications in econometrics,
one of its earliest appearances is 
in \cite{gourieroux:87}, where the
directed information functional is defined 
as a {\em Kullback causality measure}
and is derived as the limiting form 
of a likelihood ratio statistic used in a temporal 
causality hypothesis test that is closely 
related to the test we describe in 
Section~\ref{s:DI-HT}.
For the special case of first-order Markov chains
(and under some additional assumptions),
the order-$(1/n)$ convergence rate of this
statistic to the $\chi^2$ distribution
is discussed
in \cite{gourieroux:87}.

In the physics literature
directed information has also appeared
extensively 
under the name {\em transfer entropy},
and the performance of the corresponding
plug-in estimator 
is examined in the recent works
\cite{barnett:12,anderson:14}.
There, several asymptotic results
as well as non-asymptotic bounds
are described, though the main 
emphasis appears to be less on providing
rigorous proofs and more on exploring
the possible qualitative asymptotic
properties of the plug-in.

Arguably the most effective, unifying approach
to the task of understanding the behavior of the
plug-in estimator for directed information
comes from taking the point of view 
of the asymptotic analysis of maximum 
likelihood estimates in 
theoretical statistics. This is indeed
the point of view adopted in this work,
and for the proofs of our main results
we rely on two relevant sets of tools: 
One is the 
classical central limit theorem and the
law of the iterated logarithm for Markov
chains~\cite{chung:book}, and the other
is the general asymptotic theory 
of statistical inference for Markovian 
observations~\cite{billingsley:markov}.

In this vein it should be pointed out 
(as one of the anonymous reviewers 
generously outlined at length in their report)  
that the modern development of classical 
asymptotics in statistical decision theory
provides a powerful technical approach
to the task at hand, which 
can lead to what are apparently the strongest 
possible results, and accompanied by the 
sharpest intuition.
That development, pioneered 
by Le Cam, Hajek and others since the 1980s, 
cf.~\cite{lecam:86,vandervaart:book},
is built around the 
``local asymptotic normality,'' or LAN, condition.
This requires that the log-likelihood 
ratio between models with nearby parameters,
when evaluated at observations produced by
a fixed, true distribution corresponding
to one of these two parameters,
can be appropriately approximated asymptotically
by a multivariate Gaussian.
For the class of Markov models considered
in the present context, this
is not hard to verify, 
as was done, e.g., in \cite{ogata:77}.
Then, given the LAN condition, the asymptotic 
behavior of the plug-in, as well its strong
optimality properties, can be established
by applications of what are, by now, standard
results in this area; see the
local asymptotic minimax theorem
and the convolution theorem 
in~\cite[Sections~8.7,~8.9]{vandervaart:book}.

\newpage

\section{Mutual information}
\label{s:I}
%%%%%%%%%%%%%%%%%%%%%%%%%%%%%%%%%%%%%%%%%%%%%%%%%%%%%%%%%%%%%%%%%%%%%%

\subsection{Preliminaries}
\label{s:prelim}
%%%%%%%%%%%%%%%%%%%%%%%%%%%%%%%%%%%%%%%%%%%%%%%%%%%%%%%%%%%%%%%%%%%%%%

Suppose $X$ is a discrete random variable with
values in a finite set $A$, and with a distribution
described by its probability mass function,
$P_X(x)=\Pr\{X=x\}$, for $x\in A$. The entropy 
of $X$ is, $H(X)=H(P_X)
=-\sum_{x\in A}P(x)\log P(x)$,
where, throughout the paper,
`$\log$' denotes the natural
logarithm. Viewed as
a single random element, the 
joint entropy of any finite collection
of random variables $X_1^n=(X_1,X_2,\ldots,X_n)$
is defined analogously; the 
mutual information between two random
variables $X$ and $Y$ is
$I(X;Y)=H(X)+H(Y)-H(X,Y)$;
the conditional entropy $H(X|Y)=H(X,Y)-H(Y)$;
and the conditional mutual information
$I(X;Y|Z)=H(X|Z)+H(Y|Z)-H(X,Y|Z)$.
As above, we generally write 
$X_i^j=(X_i,X_{i+1},\ldots,X_j)$,
$i\leq j$, for vectors of random variables
and similarly 
$a_i^j=(a_i,a_{i+1},\ldots,a_j)\in A^{j-i+1}$,
$i\leq j$, for strings of individual symbols 
from a finite set $A$.

The joint distribution of an arbitrary number 
of discrete random variables 
is described by their joint probability mass 
function. For example,
the joint distribution of $(X,Y,Z)$ is
denoted, $P_{XYZ}(x,y,z)=\Pr\{X=x,Y=y,Z=z\}$.
We write the induced marginal distributions
in the obvious way, e.g., 
$P_{XY}(x,y)=\Pr\{X=x,Y=y\}$ and
$P_{Z}(z)=\Pr\{Z=z\}$, and the induced conditionals
are similarly denoted, e.g.,
$P_{XY|Z}(x,y|z)=\Pr\{X=y,Y=y|Z=y\}$.

\subsection{The plug-in estimator of mutual information}
\label{s:I-markov}
%%%%%%%%%%%%%%%%%%%%%%%%%%%%%%%%%%%%%%%%%%%%%%%%%%%%%%%%%%%%%%%%%%%%%%%%%

Suppose $\Xp=\{X_n\}=\{X_n\;;\;n\geq 0\}$ is a homogeneous,
first-order Markov chain on a finite alphabet $A$,
with an arbitrary 
initial distribution for $X_0$
and 
with transition matrix
$Q=(Q(a'|a)\;;\;a,a'\in A)$. 
Although it is not necessary for much of 
what follows, in order to avoid 
uninteresting technicalities we assume that 
$Q(a'|a)>0$ for all $a,a'\in A$;
see the remarks following the statement
of Theorem~\ref{thm:I-markov} regarding how 
this assumption can be relaxed.
Then $\{X_n\}$
has a unique stationary distribution $\pi$ supported
on all of $A$.

Let $\{\bar{X}_n\}$ denote the stationary version of
$\{X_n\}$, having the same transition matrix $Q$ but
with initial distribution $\bar{X}_0\sim\pi$. We wish to estimate
the mutual information rate of the Markov chain $\{X_n\}$,
$$\lim_{n\to\infty}I(X_{n};X_0^{n-1})
=\lim_{n\to\infty}I(X_{n};X_{n-1}),$$
which is easily seen to be equal to
the mutual information of the chain
in equilibrium, namely,
$$I_\pi(X_0;X_1)=I(\bar{X}_0;\bar{X}_1)=I(P_{\bar{X}_0\bar{X}_1})=
H(\bar{X}_0)+H(\bar{X}_1)-
H(\bar{X}_0,\bar{X}_1).$$

Let $\hat{P}_{X_0X_1,n}$ denote the bivariate empirical
distribution obtained from the sample $X_0^n$,
$$\hat{P}_{X_0X_1,n}(a,a')=\frac{1}{n}
\sum_{i=1}^n\IND_{\{X_{i-1}=a,X_i=a'\}},
\;\;\; a,a'\in A,$$
where $\IND_E$ denotes the indicator function of
an event $E$, which equals $1$ when $E$ occurs
and 0 otherwise. Then
{\em the plug-in estimator} $\hat{I}_n(P_{\bar{X}_0\bar{X}_1})$ 
for $I_\pi(X_0;X_1)=I(P_{\bar{X}_0\bar{X}_1})$ 
is defined as,
$$
\hat{I}_n(P_{\bar{X}_0\bar{X}_1})
=
I(\hat{P}_{X_0X_1,n})
=
H(\hat{P}_{X_0,n})
+H(\hat{P}_{X_1,n})
-H(\hat{P}_{X_0X_1,n}),
$$
where we use the same convention for the
notation of marginals, e.g., $\hat{P}_{X_0,n}(a)$,
and conditionals, e.g., $\hat{P}_{X_1|X_0,n}(a'|a)$
with those described in Section~\ref{s:prelim}.

Recall that, since all the transition probabilities 
of $\{X_n\}$ are nonzero, the bivariate chain 
$\{Z_n=(X_{n-1},X_n)\;;\;n\geq 1\}$ on $A\times A$
is ergodic, so the ergodic theorem 
\cite{chung:book}
implies that,
as $n\to\infty$, 
$\hat{P}_{X_0,n}\to \pi$, 
$\hat{P}_{X_1,n}\to \pi$,
and $\hat{P}_{X_0X_1,n}\to P_{\bar{X}_0\bar{X}_1}$,
almost surely (a.s.). Therefore,
$\hat{I}_n(P_{\bar{X}_0\bar{X}_1})$ converges to the desired value,
$I(\bar{X}_0;\bar{X}_1)$, a.s., as $n\to\infty$.
The following result describes its finer asymptotic
behavior; its proof is given in Appendix~\ref{app:I-proof}.

\begin{theorem}
\label{thm:I-markov}
Let $\Xp=\{X_n\}$ be a Markov chain with an all positive
transition matrix $Q=(Q(a'|a))$ on the finite alphabet
$A$,
and with an arbitrary 
initial distribution.
\begin{itemize}
\item[$(i)$] 
If the random variables $\{X_n\}$
are {\em not} independent, equivalently,
if $I_\pi(X_0;X_1)>0$, then,
$$\sqrt{n}[\hat{I}_n(P_{\bar{X}_0\bar{X}_1})-
I_\pi(X_0;X_1)]\weakly N(0,\sigma^2),
\;\;\;\mbox{as}\;n\to\infty,$$
where $\weakly$ denotes convergence in distribution,
$N(0,\sigma^2)$ is the zero-mean normal distribution 
with variance $\sigma^2$, and with 
$\sigma^2$ given by 
the following limit, which exists and is finite:
\be
\sigma^2=
\lim_{n\to\infty}
\frac{1}{n}
\VAR\left[\log \left(\prod_{i=1}^n
\frac{Q(\bar{X}_i|\bar{X}_{i-1})}{\pi(\bar{X}_i)}\right)\right].
\label{eq:I-cond-var}
\ee
\item[$(ii)$] 
If the random variables $\{X_n\}$
are independent, equivalently, if
$I_\pi(X_0;X_1)=0$, then,
$$2n\hat{I}_n(P_{\bar{X}_0\bar{X}_1})\weakly \chi^2\big((m-1)^2\big),
\;\;\;\mbox{as}\;n\to\infty,$$
where $\chi^2(s)$ denotes the $\chi^2$ distribution with 
$s$ degrees of freedom, and $m=|A|$ is the size of the alphabet.
\end{itemize}
\end{theorem}

\noindent
{\bf Remarks. }
\begin{enumerate}
\item
As will become evident from the proof, the restriction
that all the transition probabilities
$Q(a'|a)$ of the chain $\{X_n\}$ are positive,
is unnecessary. Indeed, for the result of part~$(i)$
it can be entirely removed, and replaced
with the minimal assumption that $\{X_n\}$
is irreducible and aperiodic. Similarly,
for part~$(ii)$ the positivity assumption can
be significantly relaxed. For example,
of Theorem~5.2 of \cite{billingsley:markov}
gives weaker conditions under which the same 
conclusions can be obtained, e.g., if we 
restrict attention to a class of ergodic
chains whose transition matrices are allowed 
to contain zero probabilities, but the
zeros always occur at the same state transitions.

\item
One of the main messages of Theorem~\ref{thm:I-markov}
is the clear dichotomy between independence and dependence:
If the random variables $\{X_n\}$ are independent,
then $I_\pi(X_0;X_1)=0$ and the plug-in estimator
$\hat{I}_n(P_{\bar{X}_0\bar{X}_1})$ converges at a rate $O(1/n)$.
On the other hand, if the $\{X_n\}$ are not independent,
then $I_\pi(X_0;X_1)$ is strictly positive 
and the plug-in estimator $\hat{I}_n(P_{\bar{X}_0\bar{X}_1})$ converges 
at the slower rate $O(1/\sqrt{n})$. 

There is a minor caveat in the above syllogism, in that
it is only valid as long as $\sigma^2$ is strictly positive;
when $\sigma^2=0$, then even if the $\{X_n\}$ are not
independent, the plug-in estimator $\hat{I}_n(P_{\bar{X}_0\bar{X}_1})$ 
converges at a rate faster than $O(1/\sqrt{n})$. But it is easy to
see, intuitively, that $\sigma^2$ is typically nonzero
when $\hat{I}_n(P_{\bar{X}_0\bar{X}_1})$ is positive. 
In the special case of chains with a uniform stationary
distribution, this is illustrated by
Proposition~\ref{prop:zero-var} below,
proved in Appendix~\ref{app:zero}.

\item
As a consequence of the proof of Theorem~\ref{thm:I-markov},
it is fairly simple to determine the exact a.s.\ rate
of convergence of the plug-in estimator under very general 
conditions. This is stated in Corollary~\ref{cor:I-LIL} below.

\item
For the proof of the second part of the theorem
we will exploit a connection between the 
problem of estimating the mutual information
$I_\pi(X_0;X_1)$ and a classical hypothesis test
for independence, as outlined in the following 
section.
\end{enumerate}

\begin{proposition}
\label{prop:zero-var}
Suppose the stationary distribution $\pi$
of the chain $\{X_n\}$ is uniform on $A$
or, equivalently, that the transition matrix
$(Q(a'|a))$ is doubly stochastic.
Then the variance $\sigma^2$ defined 
in~(\ref{eq:I-cond-var}) is zero if and only
if the $\{X_n\}$ are i.i.d.\ and each 
$X_n$ is uniformly distributed on~$A$.
\end{proposition}

The final result of this section
gives the exact pointwise rate of
convergence for the plug-in estimator;
it is established in Appendix~\ref{app:I-LIL}.

\begin{corollary}
\label{cor:I-LIL}
Let $\Xp=\{X_n\}$ be an irreducible and
aperiodic Markov chain on the 
alphabet $A$, with an arbitrary initial
distribution. Then, as $n\to\infty,$
the plug-in estimator
satisfies,
$$\hat{I}_n(P_{\bar{X}_0\bar{X}_1})=
I_\pi(X_0;X_1)+O\left(\sqrt{\frac{\log\log n}{n}}\right)
\;\;\;\mbox{a.s.}$$
\end{corollary}

\subsection{A hypothesis test for independence}
\label{s:I-HT}
%%%%%%%%%%%%%%%%%%%%%%%%%%%%%%%%%%%%%%%%%%%%%%%%%%%%%%%%%%%%%%%%%%%%%%%%%

Suppose we wish to test the null hypothesis that
the random variables $\{X_n\}$ are independent,
within the larger hypothesis that $\{X_n\}$ is 
a Markov chain with all positive transitions.
% Assume that the initial distribution
% of $X_0$ is known and 
Take, without loss 
of generality, the alphabet to be 
$A=\{1,2,\ldots,m\}$, where $m=|A|$.  
Then, we can parametrize
all possible transition matrices $Q=Q_\theta$ 
with all-positive transition probabilities, 
by an $m(m-1)$-dimensional
vector $\theta$ restricted 
an open set $\Theta\subseteq\RL^{m(m-1)}$.
Similarly, the null hypothesis is specified 
by a lower-dimensional
open set $\Phi\subseteq \RL^{m-1}$,
which is naturally embedded within $\Theta$
via a map $h:\Phi\to\Theta$.
The details of the parametrization and
the embedding are given in the proof
of Theorem~\ref{thm:I-markov} in
Appendix~\ref{app:I-proof}, but,
informally, $\Phi$ indexes
those transition matrices
$Q_{h(\phi)}$ that consist of $m$
identical rows, exactly corresponding 
to those Markov
chains that consist
of independent random variables $\{X_n\}$.

In order to test the (composite) null hypothesis
$\Phi$ within the general model $\Theta$, 
following classical statistical theory 
we employ 
a maximum-likelihood ratio test. 
Specifically, if
we define the log-likelihood $L_n(X_0^n;\theta)$
of the sample $X_0^n$ under the distribution
corresponding to $\theta$ as,
\be
L_n(X_0^n;\theta)=\log 
\big[\mbox{Pr}_\theta(X_1^n|X_0)\big]=
\log\Big(\prod_{i=1}^nQ_\theta(X_i|X_{i-1})\Big),
\label{eq:likelihood}
\ee
then the likelihood ratio test statistic is simply
the difference,
\be
\Delta_n=2\left\{
\max_{\theta\in\Theta}L_n(X_0^n;\theta)
-\max_{\phi\in\Phi}L_n(X_0^n;h(\phi))\right\}.
\label{eq:Deltan}
\ee

In terms of hypothesis testing, there are two 
important observations to be made here.
The first, is that this statistic
is {\em exactly} equal to $2n$ times
the plug-in estimator $\hat{I}_n(P_{\bar{X}_0\bar{X}_1})$;
the computation showing this is 
performed in Appendix~\ref{app:LRT}.

\begin{proposition}
\label{prop:LRT}
Under the assumptions of Theorem~\ref{thm:I-markov}
and in the notation of this section:
\ben
\Delta_n=2n\hat{I}_n(P_{\bar{X}_0\bar{X}_1}).
% \label{eq:identical}
\een
\end{proposition}

The second important thing to note is that,
under the null hypothesis, that is, 
assuming that the random variables
$\{X_n\}$ are independent,
part~$(ii)$ of Theorem~\ref{thm:I-markov}
tells us that the distribution of
$\Delta_n=2n\hat{I}_n(P_{\bar{X}_0\bar{X}_1})$
is approximately $\chi^2((m-1)^2)$, 
which does {\em not} depend on the distribution
of the data, except only through the alphabet
size $m$.  Therefore, 
we can decide whether or not the data 
$X_0^n$ offer strong enough evidence to
reject the null hypothesis by
examining the value of 
$\Delta_n=2n\hat{I}_n(P_{\bar{X}_0\bar{X}_1})$
and then computing a $p$-value based 
on this distribution.

Conversely, as we shall see in the
proof of $(ii)$ of Theorem~\ref{thm:I-markov},
the asymptotic properties of
the estimator $\hat{I}_n(P_{\bar{X}_0\bar{X}_1})$
can be deduced from general
results about the likelihood ratio $\Delta_n$.

% \newpage

\section{Directed information}
\label{s:DI}
%%%%%%%%%%%%%%%%%%%%%%%%%%%%%%%%%%%%%%%%%%%%%%%%%%%%%%%%%%%%%%%%%%%%%%

\subsection{The directed information rate of Markov chains}
%%%%%%%%%%%%%%%%%%%%%%%%%%%%%%%%%%%%%%%%%%%%%%%%%%%%%%%%%%%%%%%%%%%%%%

Let $\Xp=\{X_n\}$ and $\Yp=\{Y_n\}$ be two 
arbitrary processes with values in the finite 
alphabets $A$ and $B$, respectively. Recall 
that
the {\em directed information}
between $X_1^n$ and $Y_1^n$ is,
$$
I(X_1^n\to Y_1^n)
=H(Y_1^n)-\sum_{i=1}^nH(Y_i|Y_1^{i-1},X_1^i),$$
and that it is zero exactly when each
$Y_i$ is conditionally independent of
$X_1^i$, given its past $Y_1^{i-1}$.
The natural interpretation of this equivalence
is to say that the directed information
is zero if and only if $\Xp$ has no 
{\em causal} influence on $\Yp$.
We are interested in the problem of 
estimating the {\em directed information rate}
between $\Xp$ and $\Yp$, defined as the 
limit,
$I(\Xp\to \Yp)=\lim_{n\to\infty}
(1/n)I(X_1^n\to Y_1^n),$
whenever it exists.

If the pairs $\{(X_n,Y_n)\}$ are independent
and identically distributed,
then it is easy to see that $I(\Xp\to \Yp)$ 
simplifies to $I(X_1;Y_1)$, and the
problem of estimating it reduces to that
discussed in the Introduction.
Of course, in this case there is
nothing to discover regarding causal 
dependence.

From now on we assume that the
pair process
$\{(X_n,Y_n)\;;\;n\geq -k+1\}$ is an 
ergodic
(namely, irreducible and aperiodic)
Markov chain on the product alphabet 
$A\times B$, of memory length $k\geq 1$,
and with an arbitrary initial distribution
for $(X_{-k+1}^0,Y_{-k+1}^0)$. We write 
$\{(\bar{X}_n,\bar{Y}_n)\}$ for the stationary
version of $\{(X_n,Y_n)\}$
with $(X_{-k+1}^0,Y_{-k+1}^0)$ distributed
according to the unique invariant measure
% $\pi(a_{-k+1}^0,b_{-k+1}^0)$
of the bivariate chain, and recall that
the distribution of 
$\{(\bar{X}_n,\bar{Y}_n)\}$ can be extended
so that it is defined for all $n=\ldots,-1,0,1,\ldots$.

In this case, the following proposition 
shows that, under appropriate conditions, 
the directed information rate can be 
expressed in simpler form.

% as a functional of only 
% the $(k+1)$-dimensional distribution 
% of $\{(X_n,Y_n)\}$, so that 
% it can easily be estimated
% and a detailed analysis of the 
% corresponding estimates can be given;
% see Section~\ref{s:DI-pi}. 
% Although the results of 
% Proposition~\ref{prop:markovDI}
% have appeared, at least implicitly,
% before, we offer a short proof
% in Appendix~\ref{app:DI-prop}
% for the sake of completeness.

\begin{proposition}
\label{prop:markovDI}
Suppose 
$\{(X_n,Y_n)\}$ is an irreducible and aperiodic
Markov chain of order no larger
than $k$, with an arbitrary initial 
distribution. Then:
\begin{itemize}
\item[$(i)$]
The entropy rate $H(\Yp)$ of the univariate 
process $\Yp=\{Y_n\}$ exists and,
$$H(\Yp)
=\lim_{n\to\infty}\frac{1}{n}H(Y_1^n)
=\lim_{n\to\infty}\frac{1}{n}H(\bar{Y}_1^n).$$
\item[$(ii)$] The directed information rate
$I(\Xp\to\Yp)$ exists and it equals,
\ben
I(\Xp\to\Yp)=
\lim_{n\to\infty}\frac{1}{n}I(X_1^n\to Y_1^n)
=H(\Yp)-H(\bar{Y}_0|\bar{X}_{-k}^0,\bar{Y}_{-k}^{-1}).
% \label{eq:markovDI}
\een
% where $\{(\bar{X}_n,\bar{Y}_n)\}$ denotes the stationary
% version of the original chain.
\item[$(iii)$]
If $\Yp=\{Y_n\}$ is also a Markov chain 
of order no larger than $k$,
then $I(\Xp\to\Yp)$
further simplifies to,
\ben
I(\Xp\to\Yp)
% =I\left(P_{\bar{X}_{-k}^0\bar{Y}_0|\bar{Y}_{-k}^{-1}}\right)
=I(\bar{Y}_0;\bar{X}_{-k}^0|\bar{Y}_{-k}^{-1}).
% \label{eq:markovDI2}
\een
\end{itemize}
\end{proposition}

% \newpage

\noindent
{\bf Remarks. }
\begin{enumerate}
\item 
Throughout this section we assume that
$\{(X_n,Y_n\})$ is a Markov
chain, not necessarily stationary
(i.e., with an arbitrary initial distribution), 
with memory no larger than some fixed $k$.
For the sake of technical convenience
we will also assume
that $\{(X_n,Y_n)\}$ has
a strictly positive transition matrix $Q$, 
$$Q(a_k,b_k|a_0^{k-1},b_0^{k-1})=
\Pr\{X_n=a_k,Y_n=b_k|X_{n-k}^{n-1}=a_0^{k-1},Y_{n-k}^{n-1}=b_0^{k-1}\}>0,$$
for all $a_0^{k}\in A^{k+1}$, $b_0^{k}\in B^{k+1}.$
As discussed in the remarks following 
Theorem~\ref{thm:DI-markov}, this assumption
can be significantly relaxed.

% We denote its unique invariant distribution by
% $\pi(a_{-k+1}^0,b_{-k+1}^0)$,
% and we  write
% $\{(\bar{X}_n,\bar{Y}_n)\}$ for the stationary 
% version of the chain.
% and we generally denote the stationary 
% measure defining the distribution of
% $\{(\bar{X}_n,\bar{Y}_n)\}$ by $Q$ so that,
% e.g., $Q_{X_0^k}=P_{\bar{X}_0^k}$,
% $Q_{X_{-k}^0Y_{-k}^0}= P_{\bar{X}_{-k}^0\bar{Y}_{-k}^0}$,
% and so on.

\item
Like mutual information, the directed information rate
$I(\Xp\to\Yp)$ also admits important operational
interpretations.
For example, in the case of a stationary $k$th order Markov 
chain $\{(X_n,Y_n)\}$ such that $\{Y_n\}$ is also a $k$th
order chain, we can use the data processing property
of mutual information in the result of part~$(iii)$ 
of the proposition to see that,
$$I(\Xp\to\Yp)=
I(Y_0;X_{-k}^0|Y_{-k}^{-1})
=I(Y_0;X_{-\infty}^0|Y_{-\infty}^{-1}).$$
This quantity is zero if and only if each
$Y_i$,
given its past $Y_{-\infty}^{i-1}$,
is conditionally independent of $X_{-\infty}^i$,
confirming our original intuition that the
directed information is only zero in the absence 
of causal influence.

\item
In the case of a general stationary chain
$\{(X_n,Y_n)\}$, without assuming anything
else about the process $\{Y_n\}$, data 
processing still implies that,
$$I(Y_0;X_{-k}^0|Y_{-k}^{-1})
=I(Y_0;X_{-\infty}^0|Y_{-k}^{-1})\geq
I(Y_0;X_{-\infty}^0|Y_{-\infty}^{-1}).$$
This is zero if and only if 
$Y_0$, given only its $k$-past $Y_{-k}^{-1}$,
is conditionally independent of $X_{-\infty}^0$.
In this case the quantity
$I(Y_0;X_{-k}^0|Y_{-k}^{-1})$
is not enough to entirely characterize
the absence of causal influence from $\Xp$ to 
$\Yp$, but knowing its value nevertheless offers
some evidence for such an influence. In particular,
knowing that it is zero (or sufficiently close to zero), 
would still imply that $\Xp$ has no (or little) causal
influence on $\Yp$.

Therefore,
even if $\Yp$ is not necessarily Markovian, 
it is always of interest
to estimate
$I(\bar{Y}_0;\bar{X}_{-k}^0|\bar{Y}_{-k}^{-1})$.
Indeed, as we explain in detail in Section~\ref{s:DI-HT},
this estimation problem is intimately related to 
a likelihood-ratio hypothesis test
for the presence of causal influence.

% It is also worth noting (and will be useful below)
% that the directed information rate of interest,
% $I(\bar{Y}_0;\bar{X}_{-k}^0|\bar{Y}_{-k}^{-1})$,
% can be thought of as a functional of the 
% $(k+1)$-dimensional joint distribution
% $P_{\bar{X}_{-k}^0\bar{Y}_{-k}^0}$.
% % as in equation~(\ref{eq:markovDI2}) in 
% the statement of Proposition~\ref{prop:markovDI},
% With a slight abuse of notation we write,
% \be
% I\left(P_{\bar{X}_{-k}^0\bar{Y}_0|\bar{Y}_{-k}^{-1}}\right)
% =
% I(\bar{Y}_0;\bar{X}_{-k}^0|\bar{Y}_{-k}^{-1}),
% \label{eq:MI-DI}
% \ee
% \begin{align}
% I&\left(P_{\bar{X}_{-k}^0\bar{Y}_0|\bar{Y}_{-k}^{-1}}\right)
% 	\nonumber\\
% &=
% 	I(\bar{Y}_0;\bar{X}_{-k}^0|\bar{Y}_{-k}^{-1})
% 	\nonumber\\
% &=
% 	\sum_{a_{-k}^0\in A^{k+1},\,b_{-k}^0\in B^{k+1}}
% 	P_{\bar{X}_{-k}^0\bar{Y}_{-k}^0}
% 	(a_{-k}^0,b_{-k}^0)
% 	\log\left(\frac{
% 	P_{\bar{X}_{-k}^0\bar{Y}_0|\bar{Y}_{-k}^{-1}}(a_{-k}^0,b_0|b_{-k}^{-1})
% 	}{
% 	P_{\bar{Y}_0|\bar{Y}_{-k}^{-1}}(b_0|b_{-k}^{-1})
% 	P_{\bar{X}_{-k}^0|\bar{Y}_{-k}^{-1}}(a_{-k}^0|b_{-k}^{-1})
% 	}
% 	\right)\nonumber\\
% &\;\;\;=
% 	\sum_{a_{-k}^0\in A^{k+1},\,b_{-k}^0\in B^{k+1}}
% 	P_{\bar{X}_{-k}^0\bar{Y}_{-k}^0}
% 	(a_{-k}^0,b_{-k}^0)
% 	\log\left(\frac{
% 	P_{\bar{X}_{-k}^0\bar{Y}_{-k}^0}(a_{-k}^0,b_{-k}^0)
% 	P_{\bar{Y}_{-k}^{-1}}(b_{-k}^{-1})
% 	}{
% 	P_{\bar{X}_{-k}^0\bar{Y}_{-k}^{-1}}(a_{-k}^0,b_{-k}^{-1})
% 	P_{\bar{Y}_{-k}^0}(b_{-k}^0)
% 	}
% 	\right),
% \label{eq:DIk2}
% \end{align}
% in order to emphasize that this is a conditional mutual
% information. 
\end{enumerate}

\subsection{The plug-in estimator of directed information rate}
\label{s:DI-pi}
%%%%%%%%%%%%%%%%%%%%%%%%%%%%%%%%%%%%%%%%%%%%%%%%%%%%%%%%%%%%%%%%%%%%%%%%%

Given a sample $(X_{-k+1}^n,Y_{-k+1}^n)$ from the
joint process $\{(X_n,Y_n)\}$, 
we define the $(k+1)$-dimensional, bivariate empirical 
distribution induced on $A^{k+1}\times B^{k+1}$, as, 
\be
\hat{P}_{X_{-k}^0Y_{-k}^0,n}(a_0^k,b_0^k)=
\frac{1}{n}\sum_{i=1}^n
\IND_{\{
X^i_{i-k}
=a_0^{k},
Y^i_{i-k}
=b_0^{k}
\}},
\;\;\;a_0^k\in A^{k+1},\,b_0^k\in B^{k+1}.
\label{eq:kempirical}
\ee
Motivated by the discussion in the remarks
following Proposition~\ref{prop:markovDI},
% and recalling~(\ref{eq:MI-DI}),
we now define {\em the plug-in estimator} for the
directed information rate $I(\Xp\to\Yp)$ as,
\be
\hat{I}_n^{(k)}
(\Xp\to\Yp)
% = 
% \hat{I}_n
% \left( P_{\bar{X}_{-k}^0\bar{Y}_0|\bar{Y}_{-k}^{-1}} \right)
= 
I(\hat{Y}_0;\hat{X}_{-k}^0|\hat{Y}_{-k}^{-1}),
\;\;\;\mbox{where}\;
(\hat{X}_{-k}^0,\hat{Y}_{-k}^0)\sim
\hat{P}_{X_{-k}^0Y_{-k}^0,n}.
\label{eq:DIplugin}
\ee
% where we will be using both 
% $
% \hat{I}^{(k)}_n(\Xp\to\Yp)
% $
% and
% $
% \hat{I}_n( P_{\bar{X}_{-k}^0\bar{Y}_0|\bar{Y}_{-k}^{-1}})
% $
% alternatingly to denote 
% the plug-in estimator, since they convey different
% parts of the intuition behind its definition.

Since all the transition probabilities 
of the bivariate chain $\{(X_n,Y_n)\}$ are nonzero, 
the $(k+1)$-dimensional chain 
$\{Z_n=(X_{n-k}^n,Y_{n-k}^n)\}$
is ergodic, so the 
ergodic theorem \cite{chung:book} 
implies that the empirical distributions
$\hat{P}_{X_{-k}^0Y_{-k}^0,n}$ converge a.s., as $n\to\infty$,
to $P_{\bar{X}_{-k}^0\bar{Y}_{-k}^0}$.
And hence, the plug-in estimator
$\hat{I}_n^{(k)}(\Xp\to\Yp)$
also converges a.s.\ to the desired value,
$I(\bar{Y}_0;\bar{X}_{-k}^0|\bar{Y}_{-k}^{-1})$.
The following result
describes its finer asymptotic behavior.

\begin{theorem}
\label{thm:DI-markovY}
Let $\{(X_n,Y_n)\}$ be a Markov chain 
of memory length $k\geq 1$, with an 
all positive transition matrix $Q$ on the finite 
alphabet $A\times B$,
and with an arbitrary 
initial distribution.
Assume that the univariate process
$\{Y_n\}$ is also a Markov chain with 
memory length $k$.
\begin{itemize}
\item[$(i)$] 
If the random variables $\{X_n\}$ {\em do}
have a causal influence on the $\{Y_n\}$,
equivalently,
if $I(\Xp\to\Yp)>0$ then,
$$\sqrt{n}\Big[
\hat{I}_n^{(k)}(\Xp\to\Yp)
-I(\Xp\to \Yp)\Big]\weakly N(0,\sigma^2),
\;\;\;\mbox{as}\;n\to\infty,$$
where the variance $\sigma^2$ is given by 
the following limit, which exists and is finite:
\be
\sigma^2=
\lim_{n\to\infty}
\frac{1}{n}
\VAR\left\{
	\log\left[\prod_{i=1}^n\left(
	\frac{
	P_{\bar{X}_{-k}^0\bar{Y}_0|\bar{Y}_{-k}^{-1}}(X_{i-k}^i,Y_i|Y_{i-k}^{i-1})
	}{
	P_{\bar{Y}_0|\bar{Y}_{-k}^{-1}}(Y_i|Y_{i-k}^{i-1})
	P_{\bar{X}_{-k}^0|\bar{Y}_{-k}^{-1}}(X_{i-k}^i|Y_{i-k}^{i-1})
	}
	\right)\right]\right\}.
\label{eq:DI-cond-var}
\ee
\item[$(ii)$] 
If the random variables $\{X_n\}$ {\em do not}
have a causal influence on the $\{Y_n\}$,
equivalently,
if $I(\Xp\to\Yp)=0$ then,
\be
2n\hat{I}_n^{(k)}(\Xp\to\Yp)
\weakly \chi^2\Big(\ell^k(m^{k+1}-1)(\ell-1)\Big),
\;\;\;\mbox{as}\;n\to\infty,
\label{eq:chisq}
\ee
where $m=|A|$ and $\ell=|B|$ are the sizes of the
alphabets $A,B$, respectively.
\end{itemize}
\end{theorem}

Theorem~\ref{thm:DI-markovY}
is an immediate consequence of the following more 
general result that does not assume that $\Yp$ is
a Markov chain, combined with 
Proposition~\ref{prop:markovDI}. 
Theorem~\ref{thm:DI-markov} is
proved in Appendix~\ref{app:DI-proof}.

\newpage

\begin{theorem}
\label{thm:DI-markov}
If $\{(X_n,Y_n)\}$ is a Markov chain 
of memory length $k\geq 1$, with an 
all positive transition matrix $Q$ on the finite 
alphabet $A\times B$,
and with an arbitrary 
initial distribution, then:
\begin{itemize}
\item[$(i)$] 
If 
$I(\bar{Y}_0;\bar{X}_{-k}^0|\bar{Y}_{-k}^{-1})$
is nonzero, then with $\sigma^2$ as
in~(\ref{eq:DI-cond-var}):
$$\sqrt{n}\Big[
\hat{I}^{(k)}_n(\Xp\to\Yp)
-
I(\bar{Y}_0;\bar{X}_{-k}^0|\bar{Y}_{-k}^{-1})
\Big]\weakly N(0,\sigma^2),
\;\;\;\mbox{as}\;n\to\infty.$$
\item[$(ii)$] 
If, on the other hand,
$I(\bar{Y}_0;\bar{X}_{-k}^0|\bar{Y}_{-k}^{-1})=0,$
then the plug-in estimator 
converges
to a $\chi^2$ distribution exactly as in~(\ref{eq:chisq}).
\end{itemize}
\end{theorem}

% \newpage

\noindent
{\bf Remarks. }
\begin{enumerate}
\item
From the 
proof of Theorem~\ref{thm:DI-markov} 
it is evident that the restriction
of all-positive transition probabilities
$Q(a_k,b_k|a_{0}^{k-1},b_0^{k-1})$ for the chain 
$\{(X_n,Y_n)\}$ is unnecessary. 
The result of part~$(i)$
remains valid with this restriction
replaced with the minimal assumption that 
the pair process $\{(X_n,Y_n)\}$
is irreducible and aperiodic. 
And for part~$(ii)$
the positivity assumption can
also be significantly relaxed, 
in accordance with the discussion
around Theorem~5.2 of \cite{billingsley:markov},
particularly as long as the $k$-dimensional
version of the assumptions in Condition~5.1
is satisfied, as discussed in Remark~1
after Theorem~\ref{thm:I-markov}
earlier.

\item
An important consequence 
of Theorems~\ref{thm:DI-markovY}
and~\ref{thm:DI-markov}
is the clear dichotomy between the presence
and absence of causal influence:
If the $\{X_n\}$ have no causal influence
on the $\{Y_n\}$, 
then $I(\Xp\to\Yp)=0$ and the plug-in estimator
converges at a rate $O(1/n)$.
On the other hand, if such causal influence
does exist, then the directed information
rate $I(\Xp\to\Yp)$ is strictly positive
and the plug-in estimator 
converges at the slower rate $O(1/\sqrt{n})$. 

\item
An examination of the proof of Theorem~\ref{thm:DI-markov}
shows that, with some additional effort, it can be refined
to provide very accurate results on the a.s.\ and 
$L^1$ rates of convergence 
of the plug-in estimator, under very general 
conditions; see Corollary~\ref{cor:DI-LIL} below,
proved in Appendix~\ref{app:DI-LIL}.
In particular, in view of the 
converse result in \cite[Proposition~3]{weissman-et-al:di},
the asymptotic bound in~(\ref{eq:L1g}) implies
that the $L^1$ rate at which the plug-in converges
is optimal.

Although it is easily established that
an analogous result holds for the plug-in 
estimator of mutual information, the
corresponding proof is merely a simplification
of the (already fairly straightforward)
proof of Corollary~\ref{cor:DI-LIL}, and therefore
it was not included in the previous section.

\item
For the
proof of the $\chi^2$ convergence part 
of the theorem we will exploit an interesting
connection of this problem with a classical 
hypothesis test for conditional
independence; this is discussed
in detail in Section~\ref{s:DI-HT}.
\end{enumerate}

\begin{corollary}
\label{cor:DI-LIL}
Let $\{(X_n,Y_n)\}$ be an irreducible
and aperiodic Markov chain on $A\times B$,
of memory length $k\geq 1$, and with an 
arbitrary initial distribution.
Then, as $n\to\infty,$
the plug-in estimator
satisfies, as $n\to\infty$,
\be
\hat{I}_n^{(k)}(\Xp\to\Yp)
-
% \hat{I}_n\Big( P_{\bar{X}_{-k}^0\bar{Y}_0|\bar{Y}_{-k}^{-1}}\Big) =
I(\bar{Y}_0;\bar{X}_{-k}^0|\bar{Y}_{-k}^{-1})
&=&
O\left(\sqrt{\frac{\log\log n}{n}}\right)
\;\;\;\mbox{a.s.,}
\label{eq:LILg}\\
E\left[\big|\hat{I}_n^{(k)}(\Xp\to\Yp)
-
I(\bar{Y}_0;\bar{X}_{-k}^0|\bar{Y}_{-k}^{-1})\big|\right]
&=&
O\left(\frac{1}{\sqrt{n}}\right).
\label{eq:L1g}
\ee
If $\Yp$ is also a Markov chain of memory
no larger than $k$, then, as $n\to\infty$,
\be
\hat{I}_n^{(k)}(\Xp\to\Yp)-
% \hat{I}_n\Big( P_{\bar{X}_{-k}^0\bar{Y}_0|\bar{Y}_{-k}^{-1}}\Big)=
I(\Xp\to\Yp)
&=&
O\left(\sqrt{\frac{\log\log n}{n}}\right)
\;\;\;\mbox{a.s.}
\nonumber\\
E\left[\big|\hat{I}_n^{(k)}(\Xp\to\Yp)-
% \hat{I}_n\Big( P_{\bar{X}_{-k}^0\bar{Y}_0|\bar{Y}_{-k}^{-1}}\Big)=
I(\Xp\to\Yp)\big|\right]
&=&
O\left(\frac{1}{\sqrt{n}}\right).
\label{eq:L1}
\ee
\end{corollary}

In view of 
\cite[Proposition~3]{weissman-et-al:di},
the $L^1$ convergence rate established
in~(\ref{eq:L1}) above is optimal.

\subsection{A hypothesis test for causal influence}
\label{s:DI-HT}
%%%%%%%%%%%%%%%%%%%%%%%%%%%%%%%%%%%%%%%%%%%%%%%%%%%%%%%%%%%%%%%%%%%%%%%%%

Suppose we wish to test whether or not
the samples $\{X_n\}$ have a
causal influence on the $\{Y_n\}$. As discussed
already, in the present
context this translates
to testing the null hypothesis that
each random variable $Y_i$ is conditionally
independent of $X_{i-k}^i$ given $Y_{i-k}^{i-1}$,
within the larger hypothesis that the pair 
process $\{(X_n,Y_n)\}$ is a $k$th order 
Markov chain on $A\times B$ with all positive 
transitions.
We take, without loss of generality, the
alphabets of $\Xp$ and $\Yp$ to be 
$A=\{1,2,\ldots,m\}$
and
$B=\{1,2,\ldots,\ell\}$,
respectively.

As we describe in detail in the 
proof of Theorem~\ref{thm:DI-markov}
in Appendix~\ref{app:DI-proof},
each positive transition matrix $Q=Q_\theta$
is indexed by 
a parameter vector $\theta$ taking
values in an
$m^k\ell^k(m\ell-1)$-dimensional
open set $\Theta$.
And the null hypothesis corresponding 
to each random variable $Y_i$ being conditionally
independent of $X_{i-k}^i$ given $Y_{i-k}^{i-1}$,
is described by transition matrices $Q_\theta$
which can be decomposed as,
\be
Q_\theta(a_0,b_0| a_{-k}^{-1},b_{-k}^{-1})
=Q^x_\theta(a_0| a_{-k}^{-1},b_{-k}^{-1})
Q^y_\theta(b_{0}| b_{-k}^{-1}).
\label{eq:decomposition}
\ee
This is formally described by a lower-dimensional
parameter set $\Phi$, which can be embedded in
$\Theta$ via a map $h:\Phi\to\Theta$, such that
all induced transition matrices $Q_{h(\phi)}$
correspond to Markov chains that satisfy the
required conditional independence property~(\ref{eq:decomposition}).

In order to test the null hypothesis
$\Phi$ within the general model $\Theta$, 
we employ a likelihood ratio test. Specifically, 
we define the log-likelihood 
$L_n(X_{-k+1}^n,Y_{-k+1}^n;\theta)$
of the sample 
$(X_{-k+1}^n,Y_{-k+1}^n)$
under the distribution
corresponding to $\theta$ as,
\be
L_n(X_{-k+1}^n,Y_{-k+1}^n;\theta)=\log
\big[\mbox{Pr}_\theta(X_1^n,Y_1^n|X_{-k+1}^0,Y_{-k+1}^0)\big]=
\log\Big(\prod_{i=1}^nQ_\theta(X_i,Y_i|X_{i-k}^{i-1},Y_{i-k}^{i-1})\Big),
\label{eq:DIlikelihood}
\ee
so that the 
likelihood ratio test statistic is simply
the difference,
\be
\Delta_n=2\left\{
\max_{\theta\in\Theta}L_n(X_{-k+1}^n,Y_{-k+1}^n;\theta)
-\max_{\phi\in\Phi}L_n(X_{-k+1}^n,Y_{-k+1}^n;h(\phi))\right\}.
\label{eq:DIDeltan}
\ee

As in the earlier case discussed
in Section~\ref{s:I-HT},
there are two 
key observation to be made here.
First,
this statistic is {\em exactly} equal 
to $2n$ times the plug-in estimator.
Proposition~\ref{prop:DILRT} is 
proved in Appendix~\ref{app:DILRT}.

\begin{proposition}
\label{prop:DILRT}
Under the assumptions of Theorem~\ref{thm:DI-markov}
and in the notation of this section:
\ben
\Delta_n=2n \hat{I}_n^{(k)}
(\Xp\to\Yp).
% \label{eq:DIidentical}
\een
\end{proposition}

Recall that,
under the null hypothesis,
part~$(ii)$ of Theorem~\ref{thm:DI-markov}
tells us that the distribution of
$\Delta_n$
% =2n \hat{I}_n\Big( P_{\bar{X}_{-k}^0\bar{Y}_0|\bar{Y}_{-k}^{-1}}\Big)$
is approximately $\chi^2$ with
$\ell^k(m^{k+1}-1)(\ell-1)$
degrees of freedom. 
The second important thing to note is 
that this limiting distribution does
not depend on the actual distribution
of the samples, except through the 
alphabet sizes $m,\ell$ and the memory
length $k$. 
Therefore, following standard statistical
methodology,
we can decide whether or not the data 
offer strong enough evidence to
reject the null hypothesis by
examining the value of 
$\Delta_n$: If the $p$-value given
by the probability of the
tail $[\Delta_n,\infty)$
of the asymptotic $\chi^2$ distribution
is below a certain threshold $\alpha$,
then the causality hypothesis can be
rejected at the significance level $\alpha$.

Conversely, as we discuss in the proof
of part~$(ii)$ of Theorem~\ref{thm:DI-markov},
the asymptotic distribution of 
the plug-in estimator under the null
hypothesis, 
% $\hat{I}_n\Big( P_{\bar{X}_{-k}^0\bar{Y}_0|\bar{Y}_{-k}^{-1}}\Big)$
follows from the corresponding 
general results about the likelihood 
ratio in \cite{billingsley:markov}.

\newpage

\appendix
%%%%%%%%%%%%%%%%%%%%%%%%%%%%%%%%%%%%%%%%%%%%%%%%%%%%%%%%%%%%%%%

\section{Appendix}
%%%%%%%%%%%%%%%%%%%%%%%%%%%%%%%%%%%%%%%%%%%%%%%%%%%%%%%%%%%%%%%

\subsection{Proof of Theorem~\ref{thm:I-markov}}
\label{app:I-proof}
%%%%%%%%%%%%%%%%%%%%%%%%%%%%%%%%%%%%%%%%%%%%%%%%%%%%%%%%%%%%%%%%%%%%%%%%%
For part~$(i)$, 
first, we express
$\hat{I}_n(P_{\bar{X}_0\bar{X}_1})=I(\hat{P}_{X_0X_1,n})$ as,
\begin{align}
	\sum_{a,a'} &
	\hat{P}_{X_0X_1,n}(a,a')
	\log
	\left(\frac{\hat{P}_{X_0X_1,n}(a,a')}{
	\hat{P}_{X_0,n}(a)
	\hat{P}_{X_1,n}(a')}\right)
	\nonumber\\
&=
	\sum_{a,a'} \hat{P}_{X_0X_1,n}(a,a')
	\log\left(\frac{Q(a'|a)}{\pi(a')}\right)
	+\sum_{a,a'} \hat{P}_{X_0X_1,n}(a,a')\log 
	\left(\frac
	{\pi(a')}
	{Q(a'|a)}
	\frac{\hat{P}_{X_1|X_0,n}(a'|a)}{\hat{P}_{X_1,n}(a')}
	\right)	
	\nonumber\\
&=
	\sum_{a,a'}\left[\left(\frac{1}{n}\sum_{i=1}^n
	\IND_{\{X_{i-1}=a,X_i=a'\}}
	\right)
	\log\left(\frac{Q(a'|a)}{\pi(a')}\right)
	\right]
	\nonumber\\
&
	\;\;\;\;
	+\sum_{a,a'} \hat{P}_{X_0X_1,n}(a,a')\log 
	\left(\frac
	{\pi(a)}{\hat{P}_{X_0,n}(a)}
	\frac
	{\pi(a')}{\hat{P}_{X_1,n}(a')}
	\frac
	{\hat{P}_{X_0X_1,n}(a,a')}
	{Q(a'|a)\pi(a)}
	\right)
	\nonumber\\
&=
	\frac{1}{n}\sum_{i=1}^n
	\left[
	\sum_{a,a'}
	\left(
	\IND_{\{X_{i-1}=a,X_i=a'\}}
	\log\left(\frac{Q(a'|a)}{\pi(a')}\right)
	\right)
	\right]
	\nonumber\\
&
	\;\;\;\;
	+D(\hat{P}_{X_0X_1,n}\|P_{\bar{X}_0\bar{X}_1})
	-D(\hat{P}_{X_0,n}\|\pi)
	-D(\hat{P}_{X_1,n}\|\pi)
	\nonumber\\
&=
	\frac{1}{n}\sum_{i=1}^n
	\log\left(\frac{Q(X_i|X_{i-1})}{\pi(X_i)}\right)
	+D(\hat{P}_{X_0X_1,n}\|P_{\bar{X}_0\bar{X}_1})
	-D(\hat{P}_{X_0,n}\|\pi)
	-D(\hat{P}_{X_1,n}\|\pi),
	\label{eq:i-expand}
\end{align}
where $D(P\|Q)=\sum_{x\in A} P(x)\log[P(x)/Q(x)]$ denotes the
relative entropy between two discrete distributions
$P$ and $Q$ on the same alphabet $A$.
As mentioned earlier, the bivariate chain
$\{Z_n=(X_{n-1},X_n)\;;\;n\geq 1\}$ on $A\times A$
is ergodic, therefore it satisfies the
ergodic theorem, the central limit theorem,
and the law of the iterated logarithm;
see, e.g., \cite{chung:book}. In particular,
the law of the iterated logarithm implies that,
each of the three relative entropies above,
when multiplied by $\sqrt{n}$, converges to
zero a.s., as $n\to\infty$. Indeed, 
a Taylor expansion shows that 
$-\sqrt{n}D(\hat{P}_{X_0,n}\|\pi)$
is equal to,
\begin{align}
	\sqrt{n} \sum_{a\in A} \hat{P}_{X_0,n}(a)
&
	\log 
	\Big[1+\Big(\frac{\pi(a)}{\hat{P}_{X_0,n}(a)}-1\Big)\Big]
	\nonumber\\
&=
	\sqrt{n} \sum_{a\in A} \hat{P}_{X_0,n}(a)
	\left[
	\Big(\frac{\pi(a)}{\hat{P}_{X_0,n}(a)}-1\Big)  
	-\frac{1}{2}\Big(\frac{\pi(a)}{\hat{P}_{X_0,n}(a)}-1\Big)^2  
	\frac{1}{\xi_n(a)^2}
	\right]
	\nonumber\\
&=
	-\frac{\sqrt{n}}{2} \sum_{a\in A} \hat{P}_{X_0,n}(a)
	\left[
	\Big(\frac{\pi(a)}{\hat{P}_{X_0,n}(a)}-1\Big)^2  
	\frac{1}{\xi_n(a)^2}
	\right]
	\nonumber\\
&=
	-\frac{\sqrt{n}}{2} \sum_{a\in A}
	\left[
	\frac{1}{\hat{P}_{X_0,n}(a)\xi_n(a)^2}
	(\hat{P}_{X_0,n}(a)-\pi(a))^2
	\right],
	\nonumber
\end{align}
for some (random) $\xi_n(a)$ between 1 and $\pi(a)/\hat{P}_{X_0,n}(a)$.
By the ergodic theorem,
$(1/\hat{P}_{X_0,n}(a))$ and $(1/\xi_n(a)^2)$ are both bounded a.s., 
and from the law of the iterated logarithm we have
that $(\hat{P}_{X_0,n}(a)-\pi(a))\sqrt{n/(\log\log n)}$
is also bounded a.s. Therefore, writing
the above expression as,
\be	
	-\frac{\log \log n}{2\sqrt{n}} \sum_{a\in A}
	\left[
	\frac{1}{\hat{P}_{X_0,n}(a)\xi_n(a)^2}
	(\hat{P}_{X_0,n}(a)-\pi(a))^2\Big(\frac{n}{\log \log n}\Big)
	\right],
	\label{eq:toLIL1}
\ee
we see that each term in the sum is a.s. bounded, and,
therefore, the entire expression tends to zero, a.s.,
as $n\to\infty$, as required. 

The same argument
shows that, after multiplication
by $\sqrt{n}$, the other two relative 
entropies in~(\ref{eq:i-expand}) also 
tend to zero, a.s.,
as $n\to\infty$. Therefore,
a.s. as $n\to\infty$,
\be
\sqrt{n}[\hat{I}_n(P_{\bar{X}_0\bar{X}_1})-I_\pi(X_0;X_1)]
=
	\frac{1}{\sqrt{n}}\sum_{i=1}^n\left[
	\log\left(\frac{Q(X_i|X_{i-1})}{\pi(X_i)}\right)
	-I_\pi(X_0;X_1)\right]
+o(1),
\label{eq:toLIL2}
\ee
and the result of part~$(i)$ follows immediately
by an application of the central limit theorem
\cite[Sec.~I.16]{chung:book} to the bivariate
chain $\{Z_n\}$. The existence
of the limit in the definition of $\sigma^2$
is guaranteed by 
\cite[Theorem~3, p.~97]{chung:book}, and its
finiteness follows easily from the fact that
the alphabet is finite and 
the random variables
$\{\log \left(Q(\bar{X}_i|\bar{X}_{i-1})/\pi(\bar{X}_i)\right)\}$
are uniformly bounded.

For part~$(ii)$, first recall 
the hypothesis testing setup presented
in Section~\ref{s:I-HT}. Each
transition matrix $Q=Q_\theta$
on $A$ with positive transitions 
is indexed by a 
parameter $\theta$ in 
the following open subset of 
$\RL^{m(m-1)}$,
$$\Theta=\left\{\theta\in\RL^{m(m-1)}\;:\;\theta_{ij}>0\;
\mbox{for all}\;i,j,\;\mbox{and}\;\sum_{1\leq j\leq m-1}\theta_{ij}<1
\;\mbox{for each}\;i\right\}.$$
The entries of the transition matrix $Q_\theta$ 
corresponding to a given parameter $\theta\in\Theta$
are,
\be
Q_\theta(j|i)=\left\{
\begin{array}[c]{cc}
\theta_{ij},\;				&\text{if}\;j\neq m\\
1-\sum_{1\leq j'\leq m-1}\theta_{ij'},\;  &\text{if}\;j=m
\end{array}
\right\},
\;\;\;\mbox{for all}\;i,j.
\label{eq:parametrize}
\ee
Similarly, the null hypothesis is specified by the open set,
$$\Phi =
\left\{\phi\in\RL^{m-1}\;:\;\phi_{j}>0\;
\mbox{for all}\;j,\;\mbox{and}\;\sum_{1\leq j\leq m-1}\phi_{j}<1\right\},$$ 
which is naturally embedded within $\Theta$
via the map $h:\Phi\to\Theta$, where
$\phi\mapsto \theta=h(\phi)$, with
$\theta_{ij}=\phi_j$ for all $i,j$.

Now, recall the result of 
Proposition~\ref{prop:LRT},
stating that our
quantity of interest,
$2n\hat{I}_n(P_{\bar{X}_0\bar{X}_1})$,
is equal to twice the log-likelihood ratio
$\Delta_n$ defined in~(\ref{eq:Deltan}).
Then, the claimed convergence of
$2n\hat{I}_n(P_{\bar{X}_0\bar{X}_1})$
is exactly the result stated as
(the last part of) the conclusion of
\cite[Theorem~5.2]{billingsley:markov}.
For that, we only need to 
verify the two assumptions of that
theorem.

For the first assumption, 
we note that,
in the notation of \cite{billingsley:markov},
the size $s$ of the alphabet $s=m$, 
the dimensionality $r$ of $\Theta$ is $r=m(m-1)$,
and the dimensionality $c$ of $\Phi$ is $c=m-1$,
so that the limiting distribution of $\Delta_n$
has $r-c=(m-1)^2$ degrees of freedom.
For each $(i,j)$, $1\leq i\leq m,$ $1\leq j\leq m-1$,
the $(i,j)^{th}$
component of $h$ given by 
$h_{ij}(\phi)=\phi_j$ is 
certainly three times continuously
differentiable with respect to 
each $\phi_\ell$, $1\leq\ell\leq m-1$.
Moreover, the $m(m-1)\times(m-1)$ matrix
$K(\phi)$ with entries,
$$\big(K(\phi)\big)_{ij,\ell}
=\frac{\partial h_{ij}(\phi)}{\partial\phi_\ell}
=\frac{\partial \phi_j}{\partial\phi_\ell}
=\delta_{j\ell},$$
where $\delta_{j\ell}=\delta_{j,\ell}$ equals $1$ if $j=\ell$ and 
0 otherwise, has rank $c=m-1$ throughout $\Phi$.
This shows that the first assumption we needed,
namely, Condition~3.1 on \cite[p.~17]{billingsley:markov},
indeed is satisfied.

Similarly, for the second assumption 
let $D=A\times A$ and write $d=|D|=m^2$.
For each $(i,j)$, $1\leq i,j\leq m,$ the
component of the transition matrix $Q_\theta$ 
corresponding to $(i,j)$,
$Q_\theta(j|i)$, is certainly three times continuously
differentiable throughout $\Theta$. Moreover,
recalling the parametrization~(\ref{eq:parametrize}),
consider
the $m^2\times m(m-1)$ matrix $\clK(\theta)$ with
entries, for $1\leq i,j,\ell\leq m$, $1\leq k\leq m-1$,
$$
\big(\clK(\theta)\big)_{ij,\ell k}=
\frac{\partial Q_\theta(j|i)}{\partial \theta_{\ell k}}
=\left\{
\begin{array}[c]{cc}
\delta_{ij,\ell k},\;
	&\text{if}\;j\neq m;\\
-\delta_{i,\ell},\;
	&\text{if}\;j=m.
\end{array}
\right. 
$$
Clearly $\clK(\theta)$ has rank $r=m(m-1)$ throughout $\Theta$,
which shows that the second assumption we needed to check,
namely Condition~5.1 on \cite[p.~23]{billingsley:markov},
is also satisfied, thus completing the proof.
\qed

\subsection{Proof of Proposition~\ref{prop:zero-var}}
\label{app:zero}
%%%%%%%%%%%%%%%%%%%%%%%%%%%%%%%%%%%%%%%%%%%%%%%%%%%%%%%%%%%%%%%%%%%%%%%%%

This result is a consequence 
of \cite[Theorem~3]{kontoyiannis-jtp};
see also the discussion in \cite{yushkevich}.
To see that, observe that, under the 
assumptions of the proposition, 
$\pi(\bar{X}_i)=1/|A|$ for all $i$ and therefore,
$$\sigma^2=
\lim_{n\to\infty}
\frac{1}{n}
\VAR\left[\log \left(\prod_{i=1}^n Q(\bar{X}_i|\bar{X}_{i-1})\right)\right],
$$
which is equal to the variance $\sigma^2$
defined in \cite[eq.~(3.2)]{kontoyiannis-jtp}.
The stronger version of Theorem~3 established in 
the last section of \cite{kontoyiannis-jtp}
implies that $\sigma^2$
is only zero when there
is a constant $q>0$ and a positive 
vector $(v(a)\;;\;a\in A)$ such that,
$$Q(a'|a) = q\frac{v(a')}{v(a)},\;\;\;\mbox{for all}\;a,a'\in A.$$
But since all $Q(a'|a)$ are assumed to be strictly
positive here, we can fix an arbitrary $a\in A$ 
and sum the above expression over all $a'\in A$ to obtain
that $q/v(a)$ should be equal to 1. Since $a$
was arbitrary, this means that the vector $v(a)$
is constant over $a$, and 
the result follows.
\qed

\subsection{Proof of Corollary~\ref{cor:I-LIL}}
\label{app:I-LIL}
%%%%%%%%%%%%%%%%%%%%%%%%%%%%%%%%%%%%%%%%%%%%%%%%%%%%%%%%%%%%%%%%%%%%%%%%%
First we note that, a careful examination of the
proof of Theorem~\ref{thm:I-markov}~part~$(i)$
shows that the entire argument remains valid
for both cases $I_\pi(X_0;X_1)>0$ and 
$I_\pi(X_0;X_1)=0$, and also, without the positivity
assumption on the transition matrix, as long as 
the chain $\Xp$ is irreducible and aperiodic,
since it is a standard exercise to show that 
the bivariate chain $\{Z_n\}$ is still ergodic
in this case.

In view of the above remarks, we observe that the
computation leading to~(\ref{eq:toLIL1}) implies
that, as $n\to\infty$,
$$
\left(\sqrt{\frac{n}{\log\log n}}\right)
D(\hat{P}_{X_0,n}\|\pi) = 
O\left(\sqrt{\frac{\log\log n}{n}}\right)=o(1),
\;\;\;
\mbox{a.s.,}$$
and the same holds for each
each of the three relative entropies in~(\ref{eq:i-expand}).
Therefore, (\ref{eq:toLIL2}) becomes,
\begin{align*}
\sqrt{\frac{n}{\log\log n}}
	&
	\big[\hat{I}_n(P_{\bar{X}_0\bar{X}_1})-I_\pi(X_0;X_1)\big]\\
	&
	=
	\frac{1}{\sqrt{n\log\log n}}\sum_{i=1}^n\left[
	\log\left(\frac{Q(X_i|X_{i-1})}{\pi(X_i)}\right)
	-I_\pi(X_0;X_1)\right]
+o(1),
\end{align*}
a.s., as $n\to\infty$, and if $\sigma^2>0$
then an application of the law
of the iterated logarithm
\cite{chung:book} gives the claimed result.
Finally, we note that, in view of the 
result in \cite[Theorem~17.5.4]{meyn-tweedie:book},
the same conclusion holds in the case $\sigma^2=0$.
\qed

\subsection{Proof of Proposition~\ref{prop:LRT}}
\label{app:LRT}
%%%%%%%%%%%%%%%%%%%%%%%%%%%%%%%%%%%%%%%%%%%%%%%%%%%%%%%%%%%%%%%%%%%%%%%%%

The first maximum in the definition of $\Delta_n$,
in view of~(\ref{eq:likelihood}), can
be expressed as,
$$\max_{\theta\in\Theta}
L_n(X_0^n;\theta)
=\max_{\theta\in\Theta}
\log\Big(\prod_{i=1}^nQ_\theta(X_i|X_{i-1})\Big)
=\max_{Q}
\sum_{i=1}^n
\log\Big(
Q(X_i|X_{i-1})\Big),
$$
where the last maximum is over
all transition matrices $Q$ with
all positive entries. Therefore,
\begin{align*}
\max_{\theta\in\Theta} L_n(X_0^n;\theta)
&=
	\max_{Q} \sum_{a,a'}n\hat{P}_{X_0X_1,n}(a,a')\log(Q(a'|a))\\
&=
	-n\min_{Q}\!
	\left\{\!
	D\left(\hat{P}_{X_0X_1,n}\Big\| Q\otimes\hat{P}_{X_0,n}\right)
	-
	\sum_{a,a'}\hat{P}_{X_0X_1,n}(a,a')\log\left(
	\frac{\hat{P}_{X_0X_1,n}(a,a')}
	{\hat{P}_{X_0,n}(a)}\right)\!
	\right\},
\end{align*}
where 
$Q\otimes\hat{P}_{X,n}$ denotes the bivariate
distribution $(Q\otimes\hat{P}_{X,n})(a,a')
=
\hat{P}_{X_0,n}(a)Q(a'|a)$.
Clearly the above expression is minimized when
the relative entropy term is zero, namely, 
when $Q(a'|a)= \hat{P}_{X_0X_1,n}(a,a')/\hat{P}_{X_0,n}(a)$,
so that,
$$\max_{\theta\in\Theta}
L_n(X_0^n;\theta)
=
\sum_{a,a'}n
\hat{P}_{X_0X_1,n}(a,a')
\log\Big(
\frac{\hat{P}_{X_0X_1,n}(a,a')}{\hat{P}_{X_0,n}(a)}
\Big)
=n[
H(\hat{P}_{X_0,n})
-
H(\hat{P}_{X_0X_1,n})].
$$
A similar (and simpler) computation,
yields that the second 
maximum in~(\ref{eq:Deltan})
is,
$$\max_{\phi\in\Phi}
L_n(X_0^n;h(\phi))=
-nH(\hat{P}_{X_0,n}).$$
Combining the last two equations
with the definitions of $\Delta_n$
and $\hat{I}_n(P_{\bar{X}_0\bar{X}_1})$ 
gives the 
claimed result. 
\qed

\subsection{Proof of Proposition~\ref{prop:markovDI}}
\label{app:DI-prop}
%%%%%%%%%%%%%%%%%%%%%%%%%%%%%%%%%%%%%%%%%%%%%%%%%%%%%%%%%%%%%%%%%%%%%%%%%

The existence of the entropy rate and 
the two expressions for $H(\Yp)$ in $(i)$
can be established in several ways. 
For example, it is easy to check that 
the bivariate process $\{(X_n,Y_n)\}$ is
asymptotically mean stationary 
(AMS) \cite{GraKie:Asymptotically:1980}.
Then the univariate process $\{Y_n\}$ can be
viewed as a stationary coding of 
$\{(X_n,Y_n)\}$, and as such it is also AMS 
\cite{gray-90:book}. The results of $(i)$
then follow immediately from the
general results in \cite{gray-90:book}.

For $(ii)$, note that we can write,
\be
	\sum_{i=1}^nH(Y_i|X_1^i,Y_1^{i-1})
&=&
	\sum_{i=1}^kH(Y_i|X_1^i,Y_1^{i-1})
	+\sum_{i=k+1}^n [H(X_i,Y_i|X_1^{i-1},Y_1^{i-1})
			-H(X_i|X_1^{i-1},Y_1^{i-1})]
			\nonumber\\
&=&
	\sum_{i=1}^kH(Y_i|X_1^i,Y_1^{i-1})
	+\sum_{i=k+1}^n [H(X_i,Y_i|X_{i-k}^{i-1},Y_{i-k}^{i-1})
			-H(X_i|X_{i-k}^{i-1},Y_{i-k}^{i-1})]
			\nonumber\\
&=&
	\sum_{i=1}^k[H(Y_i|X_1^i,Y_1^{i-1})-
	H(Y_i|X_{i-k}^i,Y_{i-k}^{i-1})]
	+\sum_{i=1}^n 
	H(Y_i|X_{i-k}^i,Y_{i-k}^{i-1}).
	\label{eq:expand}
\ee
By ergodicity and the continuity of the conditional entropy
functional for finite-alphabet distributions, we have that
$H(Y_i|X_{i-k}^i,Y_{i-k}^{i-1})\to
H(\bar{Y}_0|\bar{X}_{-k}^0,\bar{Y}_{-k}^{-1})$ as $n\to\infty$.
Therefore, dividing~(\ref{eq:expand}) by $n$ and letting $n\to\infty$, 
the first term vanishes, and 
the Ces\`{a}ro averages in the second term also converge,
	$$\frac{1}{n}\sum_{i=1}^n 
	H(Y_i|X_{i-k}^i,Y_{i-k}^{i-1})\to
	H(\bar{Y}_0|\bar{X}_{-k}^0,\bar{Y}_{-k}^{-1}).$$
The result in $(ii)$ follows from
this combined with $(i)$ and the definition of the
directed information rate.

Finally if, in addition, the process $\{Y_n\}$ is itself
a Markov chain of order (no larger than) $k$, then
its entropy rate is simply, $H(\bar{Y}_0|\bar{Y}_{-k}^{-1})$,
and the result in $(iii)$ follows trivially from $(ii)$.
\qed

\subsection{Proof of Theorem~\ref{thm:DI-markov}}
\label{app:DI-proof}
%%%%%%%%%%%%%%%%%%%%%%%%%%%%%%%%%%%%%%%%%%%%%%%%%%%%%%%%%%%%%%%%%%%%%%%%%

Both parts of the theorem
will be established along the same 
lines as the proofs of the corresponding 
results in Theorem~\ref{thm:I-markov};
for that reason, some minor details in the
computations below will be omitted. 

\subsubsection{Proof of~$(i)$}
% \label{app:DI-proof}
%%%%%%%%%%%%%%%%%%%%%%%%%%%%%%%%%%%%%%%%%%%%%%%%%%%%%%%%%%%%%%%%%%%%%%%%%

For the Gaussian convergence in part~$(i)$,
recalling the definitions of the empirical
$\hat{P}_{X_{-k}^0Y_{-k}^0,n}$
and of the plug-in estimator 
in~(\ref{eq:kempirical}) and~(\ref{eq:DIplugin}),
respectively, we first express,
$\hat{I}_n^{(k)}(\Xp\to\Yp)$ as 
a mutual information,
\begin{align}
&
	\sum_{a_0^k,b_0^k}
	\hat{P}_{X_{-k}^0Y_{-k}^0,n}
	(a_0^k,b_0^k)
	\log\left(\frac{
	\hat{P}_{X_{-k}^0Y_0|Y_{-k}^{-1},n}(a_0^k,b_k|b_{0}^{k-1})
	}{
	\hat{P}_{X_{-k}^0|Y_{-k}^{-1},n}(a_0^k|b_{0}^{k-1})
	\hat{P}_{Y_0|Y_{-k}^{-1},n}(b_k|b_{0}^{k-1})
	}
	\right)\nonumber\\
&=
	\sum_{a_{0}^k,b_{0}^k}
	\hat{P}_{X_{-k}^0Y_{-k}^0,n}
	(a_{0}^k,b_{0}^k)
	\log\left(\frac{
	\hat{P}_{X_{-k}^0Y_{-k}^0,n}(a_{0}^k,b_{0}^k)
	\hat{P}_{Y_{-k}^{-1},n}(b_{0}^{k-1})
	}{
	\hat{P}_{X_{-k}^0Y_{-k}^{-1},n}(a_{0}^k,b_{0}^{k-1})
	\hat{P}_{Y_{-k}^0,n}(b_{0}^k)
	}
	\right)\nonumber\\
&=
	\sum_{a_{0}^k,b_{0}^k}
	\hat{P}_{X_{-k}^0Y_{-k}^0,n}
	(a_{0}^k,b_{0}^k)
	\log\left(
	\frac{
	P_{\bar{X}_{-k}^0\bar{Y}_0|\bar{Y}_{-k}^{-1}}(a_{0}^k,b_k|b_{0}^{k-1})
	}{
	P_{\bar{Y}_0|\bar{Y}_{-k}^{-1}}(b_k|b_{0}^{k-1})
	P_{\bar{X}_{-k}^0|\bar{Y}_{-k}^{-1}}(a_{0}^k|b_{0}^{k-1})
	}
	\right)\label{eq:sum1}\\
&\hspace{0.2in}+
	\left.\sum_{a_{0}^k,b_{0}^k}\right\{
	\hat{P}_{X_{-k}^0Y_{-k}^0,n}
	(a_{0}^k,b_{0}^k)\nonumber\\
&\hspace{0.9in}
	\left.
	\log\left(
	\frac{
	\hat{P}_{X_{-k}^0Y_{-k}^0,n}(a_{0}^k,b_{0}^k)
	\hat{P}_{Y_0^{k-1},n}(b_{0}^{k-1})
	}{
	\hat{P}_{X_{-k}^0Y_{-k}^{-1},n}(a_{0}^k,b_{0}^{k-1})
	\hat{P}_{Y_{-k}^0,n}(b_{0}^k)
	}
	\cdot
	\frac{
	P_{\bar{X}_{-k}^0\bar{Y}_{-k}^{-1}}(a_{0}^k,b_{0}^{k-1})
	P_{\bar{Y}_{-k}^0}(b_{0}^k)
	}{
	P_{\bar{X}_{-k}^0\bar{Y}_{-k}^0}(a_{0}^k,b_{0}^k)
	P_{\bar{Y}_{-k}^{-1}}(b_{0}^{k-1})
	}
	\right)
	\right\}.
	\label{eq:sum2}
\end{align}
Substituting the definition
of $\hat{P}_{X_{-k}^0Y_{-k}^{-0},n}$
in~(\ref{eq:sum1}),
simplifying, and expanding the 
logarithm in~(\ref{eq:sum2}) a sum
of four logarithms, we obtain that 
$\hat{I}^{(k)}_n(\Xp\to\Yp)$
equals,
\begin{align}
\frac{1}{n}\sum_{i=1}^n
	\log&\left(
	\frac{
	P_{\bar{X}_{-k}^0\bar{Y}_0|\bar{Y}_{-k}^{-1}}(X_{-k}^0,Y_0|Y_{-k}^{-1})
	}{
	P_{\bar{Y}_0|\bar{Y}_{-k}^{-1}}(Y_0|Y_{-k}^{-1})
	P_{\bar{X}_{-k}^0|\bar{Y}_{-k}^{-1}}(X_{-k}^0|Y_{-k}^{-1})
	}
	\right)
+D\left(\hat{P}_{X_{-k}^0Y_{-k}^0,n}\right\|\left.
P_{\bar{X}_{-k}^0\bar{Y}_{-k}^0}\right)
	\nonumber\\
&+D\left(\hat{P}_{Y_{-k}^{-1},n}\right\|\left.
P_{\bar{Y}_{-k}^{-1}}\right)
-D\left(\hat{P}_{X_{-k}^0Y_{-k}^{-1},n}\right\|\left.
P_{\bar{X}_{-k}^0\bar{Y}_{-k}^{-1}}\right)
-D\left(\hat{P}_{Y_{-k}^0,n}\right\|\left.
P_{\bar{Y}_{-k}^0}\right).
	\label{eq:di-expand}
\end{align}
We now claim that, each of the four relative entropies
above, when multiplied by $\sqrt{n}$, converges
to zero a.s.\ as $n\to\infty$. First recall that,
as stated in the beginning of Section~\ref{s:DI-pi},
the chain
$\{Z_n=(X^n_{n-k},Y_{n-k}^n)\}$
on $A^{k+1}\times B^{k+1}$
is ergodic, so we know that it 
satisfies the ergodic theorem, the central limit theorem,
and the law of the iterated logarithm
\cite{chung:book}. As we argued in the corresponding
steps in the proof of Theorem~\ref{thm:I-markov},
a quadratic Taylor expansion of the logarithm
in the definition of 
$D\left(\hat{P}_{Y_{-k}^{-1},n}\right\|\left.
P_{\bar{Y}_{-k}^{-1}}\right)$ gives,
\begin{align}
\sqrt{n}&
	D\left(\hat{P}_{Y_{-k}^{-1},n}\right\|\left.
	P_{\bar{Y}_{-k}^{-1}}\right)
	\nonumber\\
&=
% 	-\sqrt{n} \sum_{b_{-k}^{-1}\in B^k} 
% 	\hat{P}_{Y_{-k}^{-1},n}(b_{-k}^{-1})
% 	\log 
% 	\left[1+\left(\frac{P_{\bar{Y}_{-k}^{-1}}(b_{-k}^{-1})}
% 	{\hat{P}_{Y_{-k}^{-1},n}(b_{-k}^{-1})}-1\right)\right]
% 	\nonumber\\
% &=
	-\sqrt{n} \sum_{b_{0}^{k-1}\in B^k} 
	\hat{P}_{Y_{-k}^{-1},n}(b_{0}^{k-1})
	\left[
		\left(\frac{P_{\bar{Y}_{-k}^{-1}}(b_{0}^{k-1})}
		{\hat{P}_{Y_{-k}^{-1},n}(b_{0}^{k-1})}-1\right)
		-\frac{1}{2}
		\left(\frac{P_{\bar{Y}_{-k}^{-1}}(b_{0}^{k-1})}
		{\hat{P}_{Y_{-k}^{-1},n}(b_0^{k-1})}-1\right)^2
		\frac{1}{\xi_n(b_{0}^{k-1})^2}
	\right]
	\nonumber\\
&=
	\frac{\sqrt{n}}{2} \sum_{b_{0}^{k-1}\in B^k} 
	\hat{P}_{Y_{-k}^{-1},n}(b_{0}^{k-1})
		\left(\frac{P_{\bar{Y}_{-k}^{-1}}(b_{0}^{k-1})}
		{\hat{P}_{Y_{-k}^{-1},n}(b_{0}^{k-1})}-1\right)^2
		\frac{1}{\xi_n(b_{0}^{k-1})^2}
	\nonumber\\
% &=
% 	\frac{\sqrt{n}}{2} \sum_{b_{-k}^{-1}\in B^k} 
% 		\frac{1}{
% 		\hat{P}_{Y_{-k}^{-1},n}(b_{-k}^{-1})
% 		\xi_n(b_{-1}^{-k})^2}
% 		\left(
% 		\hat{P}_{Y_{-k}^{-1},n}(b_{-k}^{-1})
% 		-P_{\bar{Y}_{-k}^{-1}}(b_{-k}^{-1})
% 		\right)^2
% 		\nonumber\\
&=
	\frac{\log\log n}{2\sqrt{n}} 
	\sum_{b_{0}^{k-1}\in B^k} 
		\left[
		\frac{1}{
		\hat{P}_{Y_{-k}^{-1},n}(b_0^{k-1})
		\xi_n(b_{0}^{k-1})^2}
		\left(
		\hat{P}_{Y_{-k}^{-1},n}(b_{0}^{k-1})
		-P_{\bar{Y}_{-k}^{-1}}(b_{0}^{k-1})
		\right)^2
		\left(
		\frac{n}{\log\log n}\right)
		\right],
		\nonumber
\end{align}
for a (possibly random) $\xi_n(b_{0}^{k-1})$ 
between 1 and 
		$P_{\bar{Y}_{-k}^{-1}}(b_{0}^{k-1})/
		\hat{P}_{Y_{-k}^{-1},n}(b_{0}^{k-1})$.
Now, as in the proof of
Theorem~\ref{thm:I-markov}, the first term in the last sum above
is bounded a.s.\ by the ergodic theorem,
and the law of the iterated logarithm implies 
that the product of the next two terms,
\begin{equation}
\left(
\hat{P}_{Y_{-k}^{-1},n}(b_{0}^{k-1})
-P_{\bar{Y}_{-k}^{-1}}(b_{0}^{k-1})
\right)^2\!
\left(
\frac{n}{\log\log n}\right)
=
\left\{\!
\frac{1}{\sqrt{n\log\log n}}
\sum_{i=1}^n
\left[\IND_{\{Y_{i-k}^{i-1}=b_{0}^{k-1}
\}}
-P_{\bar{Y}_{-k}^{-1}}(b_{0}^{k-1})\right]
\right\}^2,
\label{eq:LIL-compute}
\end{equation}
is also bounded a.s. So, each summand
is a.s.\ bounded and,
therefore, the entire expression tends 
to zero, a.s., as $n\to\infty$, as claimed. 

Exactly the same argument
shows that, after being multiplied
by $\sqrt{n}$, the other three relative 
entropies in~(\ref{eq:di-expand}) also 
tend to zero, a.s.,
as $n\to\infty$, so that, a.s.,
\begin{align}
\sqrt{n}
&\Big[
\hat{I}_n^{(k)}
(\Xp\to\Yp) -
	I(\bar{Y}_0;\bar{X}_{-k}^0|\bar{Y}_{-k}^{-1})
	\Big]
	\nonumber\\
&=
\frac{1}{\sqrt{n}}\sum_{i=1}^n
	\left[
	\log\left(
	\frac{
	P_{\bar{X}_{-k}^0\bar{Y}_0|\bar{Y}_{-k}^{-1}}(X_{-k}^0,Y_0|Y_{-k}^{-1})
	}{
	P_{\bar{Y}_0|\bar{Y}_{-k}^{-1}}(Y_0|Y_{-k}^{-1})
	P_{\bar{X}_{-k}^0|\bar{Y}_{-k}^{-1}}(X_{-k}^0|Y_{-k}^{-1})
	}
	\right)
	-
	I(\bar{Y}_0;\bar{X}_{-k}^0|\bar{Y}_{-k}^{-1})
	\right]
+o(1).
	\label{eq:sum3}
\end{align}
The Gaussian convergence in part~$(i)$ follows 
by an application of the central limit theorem
\cite[Sec.~I.16]{chung:book} to the above partial
sums of a functional of the chain $\{Z_n\}$. 
The fact that the variance
in~(\ref{eq:DI-cond-var}) exists as the 
stated limit follows from 
\cite[Theorem~3, p.~97]{chung:book}, and 
the fact that it is 
finite is a direct consequence 
of the fact that
the alphabets $A$ and $B$ are finite,
and that the random variables
$\{\log(\frac{\cdots}{\cdots})\}$ being summed in~(\ref{eq:sum3})
are uniformly bounded.
\qed

\subsubsection{Proof of~$(ii)$}
% \label{app:DI-proof}
%%%%%%%%%%%%%%%%%%%%%%%%%%%%%%%%%%%%%%%%%%%%%%%%%%%%%%%%%%%%%%%%%%%%%%%%%

Recall the hypothesis testing setup of
Section~\ref{s:DI-HT}. In that notation,
the class of all possible transition 
matrices $Q=Q_\theta$ is parametrized
by an $m^k\ell^k(m\ell-1)$-dimensional
vector,
$$\theta=
\left(
\theta_{i_1,i_2,\ldots,i_k,j_1,j_2,\ldots,j_k,i',j'}
\right)=
\left(
\theta_{i_1^k,j_1^k,i',j'}
\right),$$
where $1\leq i_1,i_2,\ldots,i_k\leq m$,
$1\leq j_1,j_2,\ldots,j_k\leq \ell$,
and $(i',j')\in A\times B$,
$(i',j')\neq(m,\ell)$; we have
used the same notation as before
for strings of symbols, 
$i_1^k\in A^k$ and $j_1^k\in B^k$.
In order for each $\theta$
to correspond to a transition probability matrix
$Q_\theta$
with positive transitions,
we restrict $\theta$ to the following open subset
of $\RL^{m^k\ell^k(m\ell-1)}$,
\begin{align*}
\Theta=
	\left\{
	\theta\in(0,1)^{m^k\ell^k(m\ell-1)}
\;:\;
\sum_{(i',j')\in A\times B,(i',j')\neq (m,\ell)}
\theta_{i_1^k,j_1^k,i',j'}
<1,
\;\mbox{for all}\;
i_1^k\in A^k,j_1^k,\in B^k\right\},
\end{align*}
so that
the entries of the corresponding $Q_\theta$ are,
\begin{equation}
Q_\theta(i',j'| i_1^k,j_1^k)
=\left\{
\begin{array}[c]{ll}
\displaystyle{\theta_{i_1^k,j_1^k,i',j'},}
	&\text{if}\;(i',j')\neq (m,\ell),\\
1-
	\displaystyle{\sum_{(u,v)\neq (m,\ell)}
	\hspace{-0.15in}
	\theta_{i_1^k,j_1^k,u,v},}
	&\text{if}\;(i',j')=(m,\ell).
\end{array}
\right. 
\label{eq:DIparametrize}
\end{equation}
The null hypothesis
is described by the parameter
space $\Phi=\Gamma^x\times\Gamma^y$,
where,
\ben
\Gamma^x
&=&
	\left\{
	\gamma^x
	=
	(\gamma^x_{i_1^k,j_1^k,i'})
	\in(0,1)^{m^k\ell^k(m-1)}
	\;:\;
	\sum_{1\leq i'\leq m-1}
	\gamma^x_{i_1^k,j_1^k,i'}
	<1,
	\;\mbox{for all}\;
	i_1^k\in A^k,
	j_1^k\in B^k
	\right\},\\
\Gamma^y
&=&
	\left\{
	\gamma^y
	=
	(\gamma^y_{j_1^k,j'})
	\in(0,1)^{\ell^k(\ell-1)}
	\;:\;
	\sum_{1\leq j'\leq \ell-1}
	\gamma^y_{j_1^k,j'}
	<1,
	\;\mbox{for all}\;
	j_1^k\in B^k
	\right\}.
\een
Clearly $\Phi$ is an open set of $\RL^{\ell^k[m^k(m-1)+(\ell-1)]}$
dimensions,
which can be naturally embedded within $\Theta$
via the map $h:\Phi\to\Theta$, where
each component of $h(\phi)=h(\gamma^x,\gamma^y)$
is,
\be
h_{i_1^k,j_1^k,i',j'}(\phi)
=
\gamma^x_{i_1^k,j_1^k,i'}
\cdot\gamma^y_{j_1^k,j'},
\label{eq:embed}
\ee
for all components 
with $i'\neq m$ and $j'\neq\ell$,
with the obvious extension to the two
`edge' cases $(m,j')$ and $(i',\ell)$.

In order to establish the $\chi^2$-convergence stated 
in the theorem,
recall Proposition~\ref{prop:DILRT}
which states that $2n$ times the plug-in estimator,
$2n\hat{I}_n^{(k)}(\Xp\to\Yp)$,
equals the log-likelihood ratio $\Delta_n$ defined 
in~(\ref{eq:DIDeltan}). The claimed
result for the plug-in is an immediate
consequence of the corresponding
convergence result for $\Delta_n$ established
by Billingsley in \cite[Theorem~6.1]{billingsley:markov},
where $\Delta_n=\bar{\chi}^2_{t,t+1}(\hat{\phi})$
in the notation of \cite{billingsley:markov}.

To apply Billingsley's result we only need to 
verify its main assumption, 
specifically Condition~6.1 on
\cite[p.~33]{billingsley:markov},
which requires that, throughout $\Phi$,
the matrix $Q_{h(\phi)}$ has continuous
third order partial derivatives,
and that the matrix ${\cal L}(\phi)$
defined below has rank
$c=\ell^k[m^k(m-1)+(\ell-1)]$.

To that end, we note that,
in the notation of \cite{billingsley:markov},
the size $s$ of the alphabet $s=m\ell$, 
the memory length of the chain $t=k$,
and the full parameter space $H_t=\Theta$
has dimension $r=s^{t+1}-s^t=m^k\ell^k(m\ell-1)$. 
And since dimensionality $c$ of the parameter space
$\Phi$ for the null hypothesis is
$c=\ell^k[m^k(m-1)+(\ell-1)]$,
the limiting distribution of $\Delta_n$
has,
$$r-c
% =m^k\ell^k(m\ell-1)
% -
% \ell^k[m^k(m-1)+(\ell-1)]
=
\ell^k(m^{k+1}-1)(\ell-1)
\;\;\;\mbox{degrees of freedom}.$$

By the definitions of $\Phi$ and the map $h$
it is obvious that each component 
of the matrix $Q_{h(\phi)}$
has third-order partial derivatives with respect 
to every component of $\phi$.
Now consider the 
$s^{t+1}\times c=(m\ell)^{k+1}\times
\ell^k[m^k(m-1)+(\ell-1)]$ 
matrix $\clL(\phi)$, with entries given
by the partial derivatives of each component
$Q_{h(\phi)}(i',j'|i_1^k,j_1^k)$
of $Q_{h(\phi)}$ with respect to every component
of $\phi=(\gamma^x,\gamma^y)$. Then the row of $\clL(\phi)$
corresponding to 
$(i_1^k,j_1^k,i',j')$
consists of all,
\be
\frac{\partial{
Q_{h(\phi)}(i',j'|i_1^k,j_1^k)
}}{\partial\gamma^x_{u_1^k,v_1^k,u'}},
\label{eq:derivatives1}
\ee
followed by all,
\be
\frac{\partial{
Q_{h(\phi)}(i',j'|i_1^k,j_1^k)
}}{\partial\gamma^y_{v_1^k,v'}}.
\label{eq:derivatives2}
\ee
In view of~(\ref{eq:DIparametrize}) 
and~(\ref{eq:embed}), 
the derivatives in~(\ref{eq:derivatives1})
are equal to zero unless $i_1^k=u_1^k$ and $j_1^k=v_1^k$,
in which case,
\begin{equation*}
\frac{\partial Q_{h(\phi)}(i',j'|i_1^k,j_1^k)}
{\partial \gamma^x_{i_1^k,j_1^k,u'}}
=\left\{
\begin{array}[c]{ll}
		\displaystyle{
	\gamma^y_{j_1^k,j'},}
	&\text{if}\;
		i'=u',
		i'\neq m\;\text{and}\;j'\neq\ell,\\
		\displaystyle{
	-\gamma^y_{j_1^k,j'},}
	&\text{if}\;
		u'\neq m,
		i'= m\;\text{and}\;j'\neq\ell,\\
		\displaystyle{
	1-\sum_{v\neq \ell}
	\gamma^y_{j_1^k,v},
	}
	&\text{if}\;
		i'=u',
		i'\neq m\;\text{and}\;j'=\ell,\\
		\displaystyle{
	\sum_{v\neq \ell}
	\gamma^y_{j_1^k,v} -1,}
	&\text{if}\;
		u'\neq m,
		i'=m\;\text{and}\;j'=\ell,\\
		0,
	&\text{in all other cases.}
\end{array}
\right. 
\end{equation*}
Similarly,
the derivatives in~(\ref{eq:derivatives2})
are equal to zero unless 
$j_1^k=v_1^k$,
in which case,
\begin{equation*}
\frac{\partial Q_{h(\phi)}(i',j'|i_1^k,j_1^k)}
{\partial\gamma^y_{v_1^k,v'}}
=\left\{
\begin{array}[c]{ll}
		\displaystyle{
	\gamma^x_{i_1^k,j_1^k,i'}
	}
	&\text{if}\;
		j'=v',
		i'\neq m\;\text{and}\;j'\neq\ell,\\
		\displaystyle{
	1-\sum_{u\neq m}
	\gamma^x_{i_1^k,j_1^k,u},
		}
	&\text{if}\;
		j'=v',
		i'= m\;\text{and}\;j'\neq\ell,\\
		\displaystyle{
	-\gamma^x_{i_1^k,j_1^k,i'},
		}
	&\text{if}\;
		v'\neq\ell,
		i'\neq m\;\text{and}\;j'=\ell,\\
		\displaystyle{
	\sum_{u\neq m}
	\gamma^x_{i_1^k,j_1^k,u}-1,}
	&\text{if}\;
		v'\neq\ell,
		i'=m\;\text{and}\;j'=\ell,\\
	0,
	&\text{in all other cases.}
\end{array}
\right. 
\end{equation*}
Since all the components of all the parameters
$\gamma^x$ and $\gamma^y$ are strictly positive,
a careful (and tedious) examination of the above expressions
reveals that the matrix $\clL(\phi)$ has rank
$c=\ell^k[m^k(m-1)+(\ell-1)]$ throughout
$\Phi$.
Therefore, Condition~6.1 of \cite{billingsley:markov}
holds as claimed, thus completing the 
proof.
\qed

\subsection{Proof of Corollary~\ref{cor:DI-LIL}}
\label{app:DI-LIL}
%%%%%%%%%%%%%%%%%%%%%%%%%%%%%%%%%%%%%%%%%%%%%%%%%%%%%%%%%%%%%%%%%%%%%%%%%
In view of Proposition~\ref{prop:markovDI},
clearly it suffices to prove the first 
two assertions~(\ref{eq:LILg}) 
and~(\ref{eq:L1g}) of the corollary.
We begin by noting that, 
as long as the pair process
$\{(X_n,\Y_n)\}$ is an irreducible and aperiodic
Markov chain of order $k$, then a
standard exercise to show that the chain 
$\{Z_n=(X_{n-k}^k,Y_{n-k}^k)\}$ 
is still ergodic. Then,
an examination of the
proof of Theorem~\ref{thm:DI-markov}~part~$(i)$
reveals that the entire argument remains valid
for both cases 
$I(\bar{Y}_0;\bar{X}_{-k}^0|\bar{Y}_{-k}^{-1})>0$
and 
$I(\bar{Y}_0;\bar{X}_{-k}^0|\bar{Y}_{-k}^{-1})=0$,
and without the positivity
assumption on the transition matrix $Q$.

For the a.s.-rate in~(\ref{eq:LILg}),
we observe that the
computation leading to~(\ref{eq:LIL-compute}) implies
that, as $n\to\infty$,
$$
\left(\sqrt{\frac{n}{\log\log n}}\right)
D\left(\hat{P}_{Y_{-k}^{-1},n}\right\|\left.
P_{\bar{Y}_{-k}^{-1}}\right)=
O\left(\sqrt{\frac{\log\log n}{n}}\right)=o(1),
\;\;\;
\mbox{a.s.,}$$
and that the same bound holds for each
each of the four relative entropies in~(\ref{eq:di-expand}).
Therefore, (\ref{eq:sum3}) becomes,
\begin{align*}
&\sqrt{\frac{n}{\log\log n}}
\Big[
\hat{I}_n^{(k)}(\Xp\to\Yp)-
I(\bar{Y}_0;\bar{X}_{-k}^0|\bar{Y}_{-k}^{-1})
\Big]
	\nonumber\\
&\;\;=
\frac{1}{\sqrt{n\log\log n}}\sum_{i=1}^n
	\left[
	\log\left(
	\frac{
	P_{\bar{X}_{-k}^0\bar{Y}_0|\bar{Y}_{-k}^{-1}}(X_{-k}^0,Y_0|Y_{-k}^{-1})
	}{
	P_{\bar{Y}_0|\bar{Y}_{-k}^{-1}}(Y_0|Y_{-k}^{-1})
	P_{\bar{X}_{-k}^0|\bar{Y}_{-k}^{-1}}(X_{-k}^0|Y_{-k}^{-1})
	}
	\right)
	-
	I(\bar{Y}_0;\bar{X}_{-k}^0|\bar{Y}_{-k}^{-1})
\right]\\
&\;\;\;\;\;\;
+o(1),\;\;\;\mbox{a.s.\ as}\;n\to\infty.
\end{align*}
Then, if $\sigma^2>0$,
the law of the iterated logarithm
\cite{chung:book} implies~(\ref{eq:LILg}) as claimed,
and if $\sigma^2=0$, the same conclusion follows
by \cite[Theorem~17.5.4]{meyn-tweedie:book}.

For the $L^1$ rate in~(\ref{eq:L1g}), we first 
recall the expression for the plug-in estimator
in (\ref{eq:di-expand}) and claim that
each of the four relative entropies there
converge to zero at a rate $O(1/n)$
in $L^1$. To see this, consider 
$D(\hat{P}_{Y_{-k}^{-1},n}\| P_{\bar{Y}_{-k}^{-1}})$;
a first-order Taylor expansion for the
logarithm in its definition gives,
$$	D\left(\hat{P}_{Y_{-k}^{-1},n}\right\|\left.
	P_{\bar{Y}_{-k}^{-1}}\right)
	=\sum_{b_{0}^{k-1}}
		\hat{P}_{Y_{-k}^{-1},n}(b_0^{k-1})
		\left(
		\frac{
		\hat{P}_{Y_{-k}^{-1},n}(b_0^{k-1})
		-P_{\bar{Y}_{-k}^{-1}}(b_{0}^{k-1})
		}
		{
		P_{\bar{Y}_{-k}^{-1}}(b_{0}^{k-1})
		}
		\right)
		\left(
		\frac{1}{\zeta_n(b_0^{k-1})}
		\right),
$$
where, 
for those $b_0^{k-1}$ for which 
		$\hat{P}_{Y_{-k}^{-1},n}(b_{0}^{k-1})$
is nonzero, 
$\zeta_n(b_{0}^{k-1})$ is a (possibly random) 
constant between 1 and 
		$\hat{P}_{Y_{-k}^{-1},n}(b_{0}^{k-1})/
		P_{\bar{Y}_{-k}^{-1}}(b_{0}^{k-1})$,
while for the remaining 
$b_0^{k-1}$, 
$\zeta_n(b_{0}^{k-1})$ can be given an
arbitrary (finite) value.
Writing $S_n(b_0^{k-1})$ for the difference
		$\hat{P}_{Y_{-k}^{-1},n}(b_{0}^{k-1})-
		P_{\bar{Y}_{-k}^{-1}}(b_{0}^{k-1})$,
and $\rho_n(b_0^{k-1})=[\zeta_n(b_0^{k-1})-1]
		P_{\bar{Y}_{-k}^{-1}}(b_{0}^{k-1})/S_n(b_0^{k-1}),$
after some simple algebra we obtain,
$$
	D\left(\hat{P}_{Y_{-k}^{-1},n}\right\|\left.
	P_{\bar{Y}_{-k}^{-1}}\right)
=
	\sum_{b_{0}^{k-1}}
		S_n(b_0^{k-1})
		\left(
		\frac{
		1+S_n(b_0^{k-1})/P_{\bar{Y}_{-k}^{-1}}(b_0^{k-1})
		}
		{
		1+\rho_n(b_0^{k-1})S_n(b_0^{k-1})/P_{\bar{Y}_{-k}^{-1}}
		(b_0^{k-1})
		}
		\right),
$$
where now each $\rho_n(b_0^{k-1})\in[0,1].$
Using the simple inequality
$(1+x)/(1+\rho x)\leq 1+x(1-\rho)$, which
holds for all $x>-1$, for each $\rho\in[0,1]$,
gives,
$$
	D\left(\hat{P}_{Y_{-k}^{-1},n}\right\|\left.
	P_{\bar{Y}_{-k}^{-1}}\right)
\leq
	\sum_{b_{0}^{k-1}}
		S_n(b_0^{k-1})
		\left(1+[1-\rho_n(b_0^{k-1})]\frac{
		S_n(b_0^{k-1})}{P_{\bar{Y}_{-k}^{-1}}(b_0^{k-1})}
		\right)
\leq
	\sum_{b_{0}^{k-1}}
		\frac{
		S_n(b_0^{k-1})^2}{P_{\bar{Y}_{-k}^{-1}}(b_0^{k-1})},
$$
because each $\rho_n$ is between 0 and 1,
and each $S_n$ sums to zero by definition.
In order to show that this relative entropy
converges to zero in $L^1$ at a rate $O(1/n)$,
it suffices to show that each term in the last sum 
does. And for that it suffices to show that
for each $b_0^{k-1}$,
\be
nE[S_n(b_0^{k-1})^2]=O(1),
\;\;\;\mbox{as}\;n\to\infty.
\label{eq:L2}
\ee
But, as in~(\ref{eq:LIL-compute}),
$S_n(b_0^{k-1})$ are simply the 
centered partial sums of a functional
of the chain $\{Z_k\}$,
$$S_n(b_0^{k-1})=
\frac{1}{n}\sum_{i=1}^n\left[\IND_{\{
		Y_{i-k}^{i-1}=b_{-k}^{1}
		\}}
		-P_{\bar{Y}_{-k}^{-1}}(b_{0}^{k-1})\right],$$
so the result of
\cite[Theorem~3, p.~97]{chung:book}
tells us that actually,
$nE[S_n(b_0^{k-1})^2]$ converges to 
a finite constant as $n\to\infty$. 
This implies~(\ref{eq:L2}), and establishes that,
$$
	E\left|D\left(\hat{P}_{Y_{-k}^{-1},n}\right\|\left.
	P_{\bar{Y}_{-k}^{-1}}\right)\right|=
	O\Big(\frac{1}{n}\Big),
	\;\;\;\mbox{as}\;n\to\infty.
$$

The exact same argument shows that the
same result also holds for the other three
relative entropies in~(\ref{eq:di-expand}).
Therefore, in order to establish the
required result in~(\ref{eq:L1g})
it suffices to show that,
\be
E\left|
\frac{1}{n}\sum_{i=1}^n
	\log\left(
	\frac{
	P_{\bar{X}_{-k}^0\bar{Y}_0|\bar{Y}_{-k}^{-1}}(X_{-k}^0,Y_0|Y_{-k}^{-1})
	}{
	P_{\bar{Y}_0|\bar{Y}_{-k}^{-1}}(Y_0|Y_{-k}^{-1})
	P_{\bar{X}_{-k}^0|\bar{Y}_{-k}^{-1}}(X_{-k}^0|Y_{-k}^{-1})
	}
	\right)-
	I(\bar{Y}_0;\bar{X}_{-k}^0|\bar{Y}_{-k}^{-1})\right|
	=O\Big(\frac{1}{\sqrt{n}}\Big).
\label{eq:last}
\ee
But, once again, this can be seen as the $L^1$ norm of the
centered partial sums of a functional of the chain
$\{Z_n\}$. Therefore, H\"{o}lder's inequality combined with 
the result of \cite[Theorem~3, p.~97]{chung:book},
imply exactly~(\ref{eq:last}), establishing~(\ref{eq:L1g})
and completing the proof.
\qed

\subsection{Proof of Proposition~\ref{prop:DILRT}}
\label{app:DILRT}
%%%%%%%%%%%%%%%%%%%%%%%%%%%%%%%%%%%%%%%%%%%%%%%%%%%%%%%%%%%%%%%%%%%%%%%%%

We proceed along the same line as in the 
proof of Proposition~\ref{prop:LRT}.
Recalling~(\ref{eq:DIlikelihood}), the
first maximum in the definition of $\Delta_n$ is,
$$\max_{\theta\in\Theta}
L_n(X_{-k+1}^n,Y_{-k+1}^n;\theta)
=\max_{Q}
\sum_{i=1}^n
\log\Big(
Q(X_i,Y_i|X_{i-k}^{i-1},Y_{i-k}^{i-1})\Big),
$$
where the last maximization
is over all transition matrices $Q$ 
with all positive entries, so that,
\begin{align*}
\max_{\theta\in\Theta}
&
	\;L_n(X_{-k+1}^n,Y_{-k+1}^n;\theta)\\
&=
	\max_{Q} \sum_{a_0^k,b_0^k}
	n
	\hat{P}_{X_{-k}^0Y_{-k}^0,n}(a_{0}^k,b_0^k)
	\log(Q(a_k,b_k|a_0^{k-1},b_0^{k-1}))\\
&=
	-n\min_{Q} 
	\mbox{\huge $\Big\{$ }
	D\left(
	\hat{P}_{X_{-k}^0Y_{-k}^0,n}(a_{0}^k,b_0^k)
	\Big\| 
	Q\otimes\hat{P}_{X_{-k}^{-1}Y_{-k}^{-1},n}\right)\\
&
	\hspace{0.9in}
	-
	\sum_{a_0^k,b_0^k}
	\hat{P}_{X_{-k}^0Y_{-k}^0,n}(a_{0}^k,b_0^k)
	\log\left(\frac{
	\hat{P}_{X_{-k}^0Y_{-k}^0,n}(a_{0}^k,b_0^k)
	}
	{
	\hat{P}_{X_{-k}^{-1}Y_{-k}^{-1},n}(a_{0}^{k-1},b_0^{k-1})
	}\right)
	\mbox{\huge $\Big\}$ },
\end{align*}
where the distribution,
$$(Q\otimes\hat{P}_{X_{-k}^{-1}Y_{-k}^{-1},n})(a_0^k,b_0^k)
=
\hat{P}_{X_{-k}^{-1}Y_{-k}^{-1},n}(a_0^{k-1},b_0^{k-1})
Q(a_k,b_k|a_0^{k-1},b_0^{k-1}),$$
for all 
$a_0^k\in A^{k+1}$,
$b_0^k\in B^{k+1}$.
The above minimum is obviously achieved
by making the relative entropy equal to zero,
that is, by taking,
$$Q(a_k,b_k|a_0^{k-1},b_0^{k-1})
=
\hat{P}_{X_0Y_0|X_{-k}^{-1}Y_{-k}^{-1},n}
	(a_k,b_k|a_0^{k-1},b_0^{k-1}),$$
so that,
\be
\max_{\theta\in\Theta}
L_n(X_{-k+1}^n,Y_{-k+1}^n;\theta)
=n\left[
H\left(
\hat{X}_{-k}^{-1},\hat{Y}_{-k}^{-1}
\right)
-
H\left(
\hat{X}_{-k}^0,\hat{Y}_{-k}^0
\right)\right],
\label{eq:firstmax}
\ee
where, $(\hat{X}_{-k}^0,\hat{Y}_{-k}^0)\sim
\hat{P}_{X_{-k}^0Y_{-k}^0,n}$.

The computation for the second 
maximum in~(\ref{eq:DIDeltan})
is a little more involved, as it
reduces to two different maximizations.
But because both of these are very similar 
to the one just computed, we will give
an outline of the steps involved without
providing all the details.
Recall the log-likelihood expression in~(\ref{eq:DIlikelihood})
and that, under the null, $Q$
admits the decomposition in~(\ref{eq:decomposition}).
We have:
\begin{align}
\max_{\phi\in\Phi}&\;
	L_n(X_{-k+1}^n,Y_{-k+1}^n;\theta)
	\nonumber\\
&=
	\max_{Q^x,Q^y}
	\sum_{i=1}^n
	\log\Big(
	Q^x(X_i|X_{i-k}^{i-1},Y_{i-k}^{i-1})
	Q^y(Y_i|Y_{i-k}^{i-1})
	\Big)
	\nonumber\\
&=
	\max_{Q^x}
	\sum_{i=1}^n
	\log\Big(
	Q^x(X_i|X_{i-k}^{i-1},Y_{i-k}^{i-1})
	\Big)
	+
	\max_{Q^y}
	\sum_{i=1}^n
	\log\Big(
	Q^y(Y_i|Y_{i-k}^{i-1})
	\Big)
	\nonumber\\
&=
	\max_{Q^x} \sum_{a_0^k,b_0^{k-1}}
	n
	\hat{P}_{X_{-k}^0Y_{-k}^{-1},n}(a_{0}^k,b_0^{k-1})
	\log(Q^x(a_k|a_0^{k-1},b_0^{k-1}))
	\nonumber\\
&
	\;\;\;+
	\max_{Q^y} \sum_{b_0^k}
	n
	\hat{P}_{Y_{-k}^0,n}(b_0^k)
	\log(Q^y(b_k|b_0^{k-1}))
	\nonumber\\
&=
	n\sum_{a_0^k,b_0^{k-1}}
	\hat{P}_{X_{-k}^0Y_{-k}^{-1},n}(a_{0}^k,b_0^{k-1})
	\log\left(\frac{
	\hat{P}_{X_{-k}^0Y_{-k}^{-1},n}(a_{0}^k,b_0^{k-1})
	}
	{
	\hat{P}_{X_{-k}^{-1}Y_{-k}^{-1},n}(a_{0}^{k-1},b_0^{k-1})
	}\right)
	\nonumber\\
&
	\;\;\;+
	n\sum_{b_0^{k-1}}
	\hat{P}_{Y_{-k}^0,n}(b_0^k)
	\log\left(\frac{
	\hat{P}_{Y_{-k}^0,n}(b_{0}^k)
	}
	{
	\hat{P}_{Y_{-k}^{-1},n}(b_0^{k-1})
	}\right)\nonumber\\
&=
	n\left[-H\left(\hat{X}_{-k}^0,\hat{Y}_{-k}^{-1}\right)
	+H\left(\hat{X}_{-k}^{-1},\hat{Y}_{-k}^{-1}\right)
	-H\left(\hat{Y}_{-k}^0\right)
	+H\left(\hat{Y}_{-k}^{-1}\right)\right].
	\label{eq:secondmax}
\end{align}
Combining~(\ref{eq:firstmax}) and~(\ref{eq:secondmax})
and using the chain rule,
\ben
\Delta_n
&=&
	2\left\{
	\max_{\theta\in\Theta}L_n(X_{-k+1}^n,Y_{-k+1}^n;\theta)
	-\max_{\phi\in\Phi}L_n(X_{-k+1}^n,Y_{-k+1}^n;h(\phi))\right\}\\
&=&
	2n\left[
	H\Big(\hat{Y}_{0}\big|\hat{Y}_{-k}^{-1}\Big)
	-H\Big(\hat{Y}_{0}\big|\hat{X}_{-k}^0,\hat{Y}_{-k}^{-1}\Big)
	\right],
\een
which, recalling the definition of
$\hat{I}_n^{(k)}(\Xp\to\Yp)$,
is precisely the claimed result.
\qed

\newpage

\section*{Acknowledgments}
%%%%%%%%%%%%%%%%%%%%%%%%%%%%%%%%%%%%%%%%%%%%%%%%%%%%%%%%%%%%%%%

We wish to thank Wojtek Szpankowski 
and Vasiliki Gioutlaki for interesting discussions
during the preliminary phase of this work.
We are also grateful to the three anonymous
reviewers for the their very detailed,
insightful and useful comments.

\bibliographystyle{plain}

% \bibliography{../../latex/ik}

\def\cprime{$'$}

%%%%%%%%%%%%%%%%%%%%%%%%%%%%%%%%%%%%%%%%%%%%%%%%%%%%%%%%%%%%%%%

\end{document}